\documentclass[journal]{IEEEtran}
\ifCLASSINFOpdf
\else
\fi
\usepackage{multirow} 
\usepackage{booktabs}
\usepackage{makecell}
\usepackage{algorithm}
\usepackage{algorithmic}
\usepackage{cite}
\usepackage{graphicx}
\usepackage{url}

\graphicspath{{./figures/}}

\usepackage{amsmath}
\ifCLASSOPTIONcompsoc
\usepackage[caption=false,font=normalsize,labelfont=sf,textfont=sf]{subfig}
\else
\usepackage[caption=false,font=footnotesize]{subfig}
\fi

\begin{document}
	%
	\title{Towards On-Device Federated Learning: A Direct Acyclic Graph-based Blockchain Approach}
	%
	%
	%
	
	
	\author{\IEEEauthorblockN{Mingrui~Cao, Long Zhang, and Bin Cao$^*$,~\IEEEmembership{Member,~IEEE}}
		\thanks{This work was supported in part by the National Natural Science Foundation of China under Grant 61701059. \textit{(Corresponding author: Bin Cao)}

			M. Cao (e-mail: leo201313@foxmail.com) and B. Cao (e-mail: caobin@bupt.edu.cn) are with the State Key Laboratory of Networking and Switching Technology, Beijing University of Posts and Telecommunications, Beijing 100876, China. M. Cao is also with University of Electronic Science and Technology of China, Chengdu 611731, China.
			
			L. Zhang (e-mail: zhanglong3211@yeah.net) is with the National Key Lab on Communications, University of Electronic Science and Technology of China, Chengdu 611731, China.
			
	}}

	\maketitle
	\begin{abstract}

		Due to the distributed characteristics of Federated Learning (FL), the vulnerability of global model and coordination of devices are the main obstacle. As a promising solution of decentralization, scalability and security, leveraging blockchain in FL has attracted much attention in recent years. However, the traditional consensus mechanisms designed for blockchain like Proof of Work (PoW) would cause extreme resource consumption, which reduces the efficiency of FL greatly, especially when the participating devices are wireless and resource-limited. In order to address device asynchrony and anomaly detection in FL while avoiding the extra resource consumption caused by blockchain, this paper introduces a framework for empowering FL using Direct Acyclic Graph (DAG)-based blockchain systematically (DAG-FL). Accordingly, DAG-FL is first introduced from a three-layer architecture in details, and then two algorithms \textit{DAG-FL Controlling} and \textit{DAG-FL Updating} are designed running on different nodes to elaborate the operation of DAG-FL consensus mechanism. After that, a Poisson process model is formulated to discuss that how to set deployment parameters to maintain DAG-FL stably in different federated learning tasks. The extensive simulations and experiments show that DAG-FL can achieve better performance in terms of training efficiency and model accuracy compared with the typical existing on-device federated learning systems as the benchmarks.

	\end{abstract}
	

	%
	\IEEEpeerreviewmaketitle

	\section{Introduction}

	\IEEEPARstart{I}{n} order to solve the privacy problem caused by the data island and make the best use of distributed data on various devices, Federated Learning (FL) has recently drawn much attention, which is a distributed machine learning framework, and participants in FL transfer and communicate the model parameters without revealing user privacy to use their own data to establish machine learning models [\citen{mcmahan2017communication}, \citen{mcmahan2017federated}]. For wireless scenarios, on-device FL is one of the most typical applications where participating nodes of FL are numerous mobile devices under a wireless network [\citen{kim2019blockchained}, \citen{hard2018federated}]. Meanwhile, with the advent of 5G era, mobile devices would have sufficient communication bandwidth which makes it possible to establish an efficient FL system on mobile devices.


	\par Although FL is widely considered to be a feasible way to enhance privacy and security in 5G wireless networks, it still faces many challenges during deployment \cite{li2020federated}. The main two points are as follows.
	
	
	\begin{itemize}
		\item \textit{\textbf{Device asynchrony}}: Various nodes have different resources for FL in terms of computing, communication, caching, battery power, data, and training time, which would result in heterogeneity and it is a natural characteristic especially in wireless networks \cite{li2020federated}. As a result, due to the limited capacity and ability of node, network and system, it is hard to coordinate FL process perfectly generating the device asynchrony \cite{9163027}. To this end, in the traditional centralized and synchronous FL system like Google FL proposed in 2017 \cite{mcmahan2017communication}, a single node must wait for other nodes to complete their tasks and then enter the next round together after completing its own training task. However, this manner might generate the deteriorated cost incurred by the bottleneck node obviously, in which the worst case is that if a node shuts down during training, it may let a round of FL be invalid completely [\citen{xie2019asynchronous}, \citen{lian2018asynchronous}].

		
		\item \textit{\textbf{Anomaly detection}}: 
		Due to privacy concerns, the local data set and local operation process of a node are invisible to others, which makes FL suffer from abnormal actions of nodes easily. Especially considering massive participating nodes, the challenge of FL is to detect abnormal nodes and avoid adverse effects as much as possible. Abnormal nodes will reduce the overall efficiency of FL system as well as model accuracy by uploading abnormal parameters during the FL process, the machine learning model built by all nodes together would be very vulnerable, and thus anomaly detection in FL is necessary [\citen{bagdasaryan2020backdoor}, \citen{xie2019slsgd}, \citen{zhang2019poisoning}].
	\end{itemize} 

	\par To this end, some researches focus on asynchronous FL framework to solve device asynchrony, and anomaly detection strategy to mitigate the impact of abnormal nodes in FL system. Recently, considering the asynchronous system and security protection, blockchain becomes a natural design that adopted in FL \cite{chen2020pervasive}, the reasons are twofold. (1) Nodes can announce local model immediately without any asynchrony requirement. (2) Blockchain miners can collect and validate model parameters to encourage normal action while avoiding anomaly.

	\par Although recent blockchained FL works [\citen{kim2019blockchained}, \citen{majeed2019flchain}, \citen{salah2019blockchain}] have achieved some progress and advantages to address the mentioned challenges, there are still some problems remained that have not been investigated thoroughly. 
	First, although these works basically follow the synchronous framework of Google FL, they allow nodes not to wait for others. Thus a pseudo-asynchronous FL system is established, in which the innovation is to use miners to replace the central servers. However, due to the synchrony, the mobile device acted as a miner should be associated with each other, which obeys the decentralization of blockchain while declining the performance of FL in terms of delay, convergence, accuracy and etc. Second, to maintain the blockchain operation, most works use PoW \cite{nakamoto2019bitcoin} which allows node acted as a miner consuming an amount of computing resource for consensus achievement. Meanwhile, in order to detect abnormal nodes, the miner should also verify the correctness of uploaded model parameters. As a result, the training/learning efficiency might not meet the expectation well caused by the blockchain cost since the overall resource is limited, especially for on-device FL under wireless network.

	\par Through the above observations, our concerned issue is whether blockchain is available to establish an efficient asynchronous on-device FL system without introducing too much extra resource consumption. Owing to the evolution of consensus mechanism, we notice that Directed Acyclic Graph (DAG) ledger technology promotes blockchain from synchronous to asynchronous bookkeeping using voting consensus mechanism without mining \cite{li2020direct}. Inspired by these, we propose an asynchronous DAG empowered FL for purpose of efficiency and immunity, referred as DAG-FL. 
	The main contributions of this work are illustrated as follows.
	\begin{itemize}
		\item To the best of our knowledge, DAG-FL is the first DAG based FL framework forming an asynchronous updating to well solve problems of device asynchrony and abnormal nodes under wireless network.
		\item We formulate a theoretical model to analyze and discuss how to keep DAG-FL working stably and autonomously, which is helpful to understand the operation of DAG-FL.
		\item We conduct extensive experiments in simulation computer system and testbed deployment to evaluate the performance of DAG-FL compared with Google, Asynchronous, and Block FL systems to discuss the improvement and feasibility of the proposed DAG-FL.
		
	\end{itemize}

	The rest of this paper is organized as follows. In Section \uppercase\expandafter{\romannumeral2} we provide some basic concepts in DAG-FL. Section \uppercase\expandafter{\romannumeral3} proposes DAG-FL in details. Furthermore, Section \uppercase\expandafter{\romannumeral4} analyzes two details of deploying DAG-FL and discusses how to make DAG-FL work properly. Then, Section \uppercase\expandafter{\romannumeral5} analyzes the performance of DAG-FL in the case of large-scale nodes through simulation experiments, and implements DAG-FL on a real testbed with a small number of nodes. As future work, we discuss the expansion of DAG-FL in Section \uppercase\expandafter{\romannumeral6}. In Section \uppercase\expandafter{\romannumeral7}, we will briefly discuss about some related works of existing FL systems from synchronous, pseudo-asynchronous, asynchronous, and blockchained aspects. Finally, Section \uppercase\expandafter{\romannumeral8} gives a summary.

	
	\section{Preliminaries}
	\par In order to elaborate the proposed DAG-FL, the basic concepts of FL and blockchain involved in DAG-FL are introduced briefly in this section.
	
	\subsection{Federated Learning}
	
	\par
\begin{figure}[tbp]
		\centering
		\subfloat[Synchronous FL]{\includegraphics[width=3in]{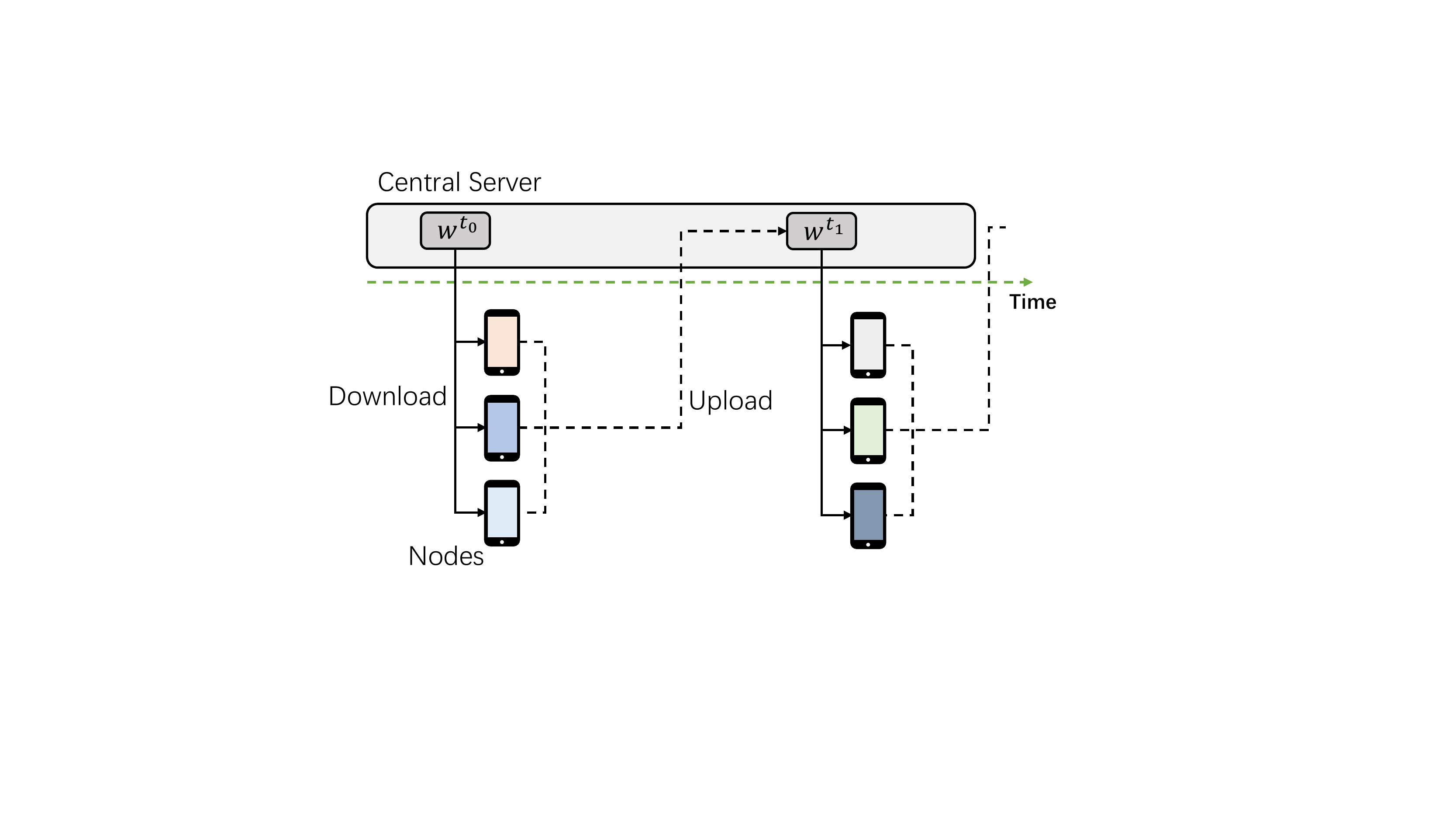}\label{p_a}}\\
		\subfloat[Pseudo-asynchronous FL]{\includegraphics[width=3in]{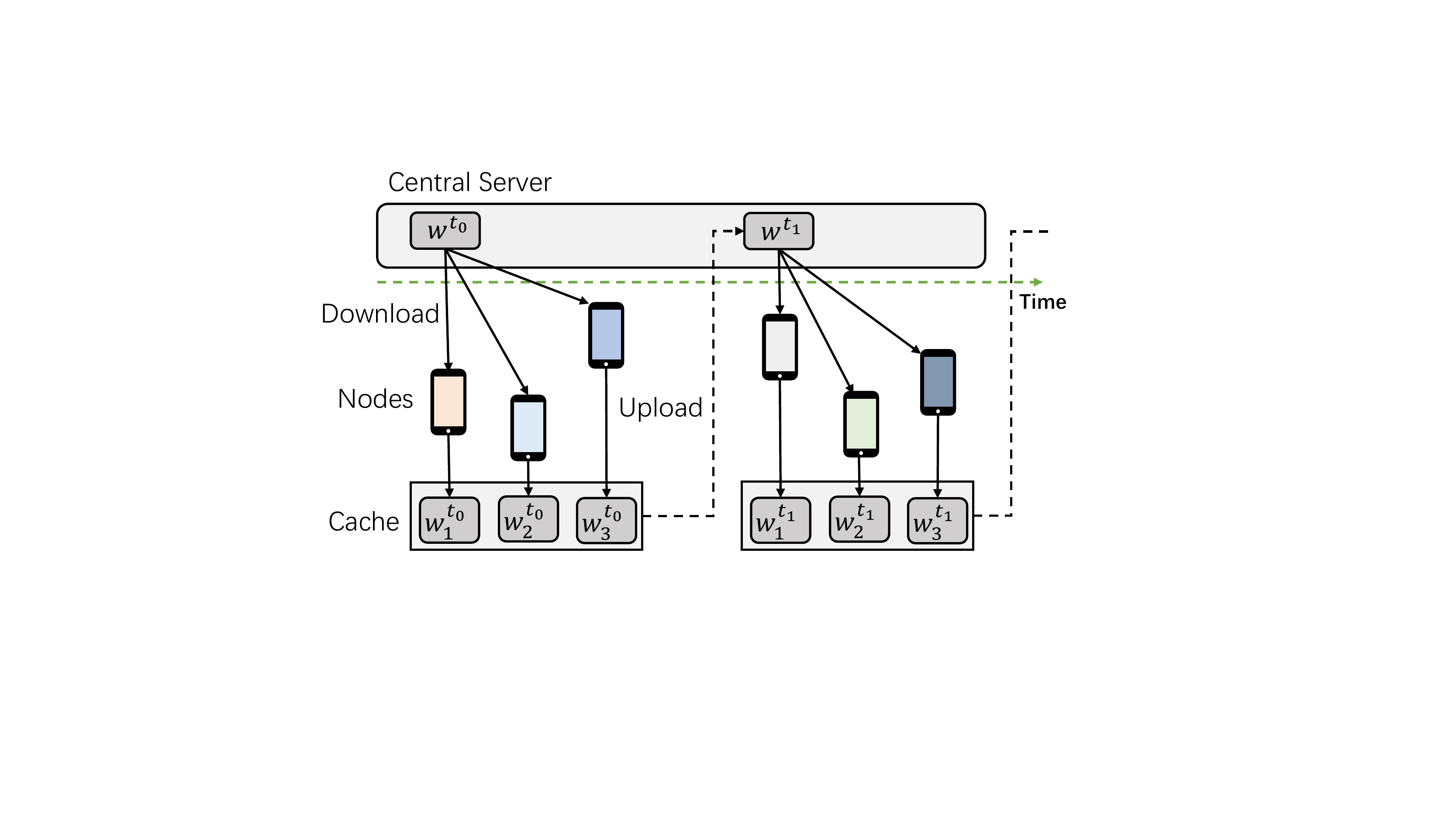}\label{p_b}}\\
		\subfloat[Asynchronous FL]{\includegraphics[width=3in]{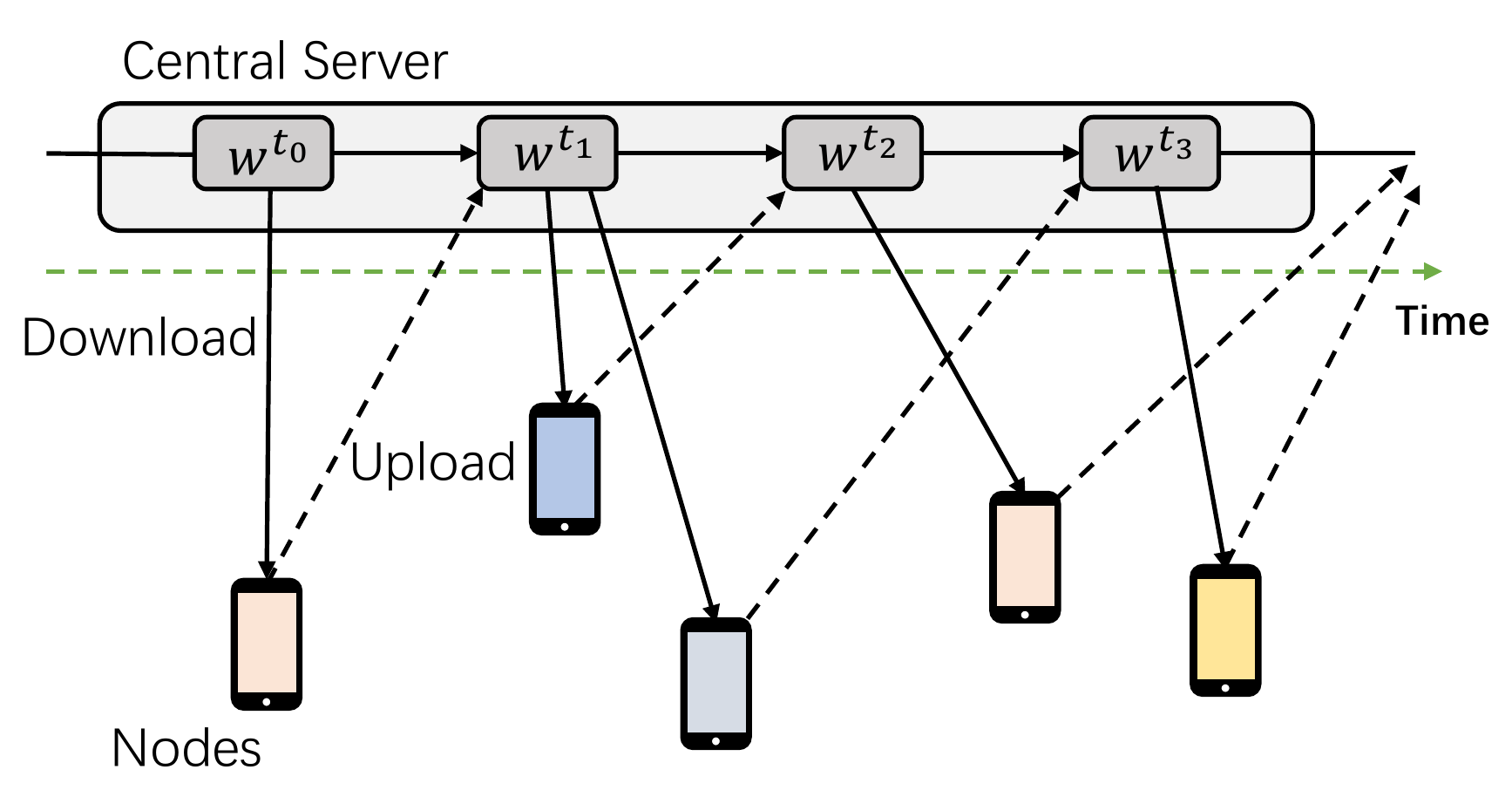}\label{p_c}}\\
		\caption{Three kinds of FL.}
		\label{pre}
	\end{figure}

 FL is a distributed machine learning method to utilize data on different devices without privacy leaking, and it is proposed by Google in 2017 officially \cite{mcmahan2017federated}. Traditional FL like Google FL is composed of a central server and numerous nodes as shown in Fig. \ref{pre}.(a), where the global model is maintained on central server and FL iterations are performed on nodes. In every round of FL, the central server selects several idle nodes to assign FL tasks. The selected nodes then begin to run FL iterations. To complete an FL iteration, the node first downloads the current global model from the central server. Second, the node trains the global model using local data to get a local model and upload it to the central server. Finally, when the central server collects all local models uploaded by selected nodes, it runs the \textit{FederatedAveraging} algorithm \cite{mcmahan2017communication} that combines all the local models on averaging to get an aggregated new global model. This global model updating method is a typical synchronous manner, and the traditional FL like Google FL is so-called synchronous FL.
	
	By uploading trained local models to a cache of the central server immediately, the pseudo-asynchronous FL \cite{9093123} is constructed as shown in Fig. \ref{pre}.(b). Nodes in pseudo-asynchronous can download global model from the central server and train it with local data whenever it is in idle state. The central server will update the global model regularly by aggregating the newly uploaded local models in the cache. To accelerate the system efficiency, pseudo-asynchronous FL should set a small interval between global model updating, but this would cause that data on bottleneck nodes could never be utilized and results in a final target model with less prediction accuracy.

	In fact, due to device asynchrony \cite{li2020federated}, neither synchronous nor pseudo-asynchronous FL could fit the on-device FL scenarios well, and thus asynchronous FL has been studied recently. In asynchronous FL, it only needs two steps to complete an iteration of FL. First, any node can download the global model from the central server whenever it is idle, and train the global model to get a local model. Then, the node uploads the local model back to the central server, and the central server updates the global model immediately whenever the local model is collected \cite{xie2019asynchronous} as shown in Fig. \ref{pre}.(c). Furthermore, asynchronous FL allows global models to be updated by local models trained from the latest and outdated global models.

	\subsection{Blockchain and DAG ledger technology}
	
	\par Blockchain is a peer-to-peer (P2P) distributed ledger technology for storing information securely and immutably \cite{taylor2020systematic}. Usually, the traditional blockchain is based on Proof of X (the most famous one is Proof of Work \cite{nakamoto2019bitcoin}), which consumes much resource on solving hash problems to decide which block to add to the chain next while slowing down the access rate of new blocks to avoid fork problem \cite{cao2020performance}.  Moreover, the  RAFT-based blockchain does not tolerate the existence of malicious nodes and PBFT-blockchain meets the challenge of communication overhead increasing with the participating nodes exponentially \cite{xu2020raft}.

	
	\par In order to break the above limitations, DAG-based blockchain (DAG ledger) is proposed to promote the synchronous blockchain to asynchronous bookkeeping. The principle of DAG-based blockchain is to attach the new transactions in a forking topology \cite{li2020direct} without maintaining a main chain, and thus any new arrival transaction can be recorded in blockchain immediately without any coordination \cite{cao2019internet}. The consensus used in DAG-based blockchain can be treated as a voting mechanism, which requires nodes to validate and approve some early published transactions before publishing their owns. The interval between publishing time of a transaction and current time is called staleness. Transactions with good staleness that are not approved yet on DAG are usually called as tips \cite{popov2016tangle}. To publish a transaction on DAG, one node needs to go through three stages. In the first stage, the node selects some tips according to some algorithms or just randomly. In the second stage, the node validates the authentication and correctness of selected tips. In the final stage, a new transaction, composed of essential information and approvals to selected tips, is constructed and published on  DAG. Through these three stages, the votes are stored in the published transactions, and unidirectional connections among transactions are built forming the DAG architecture through the approval relationship.

	\section{DAG-FL}
	In this section, we first provide the overview of DAG-FL, and then introduce its operation process.
	
	\subsection{Asynchronous Architecture}
\begin{figure}[!t]
		\centering
		\includegraphics[width=3in]{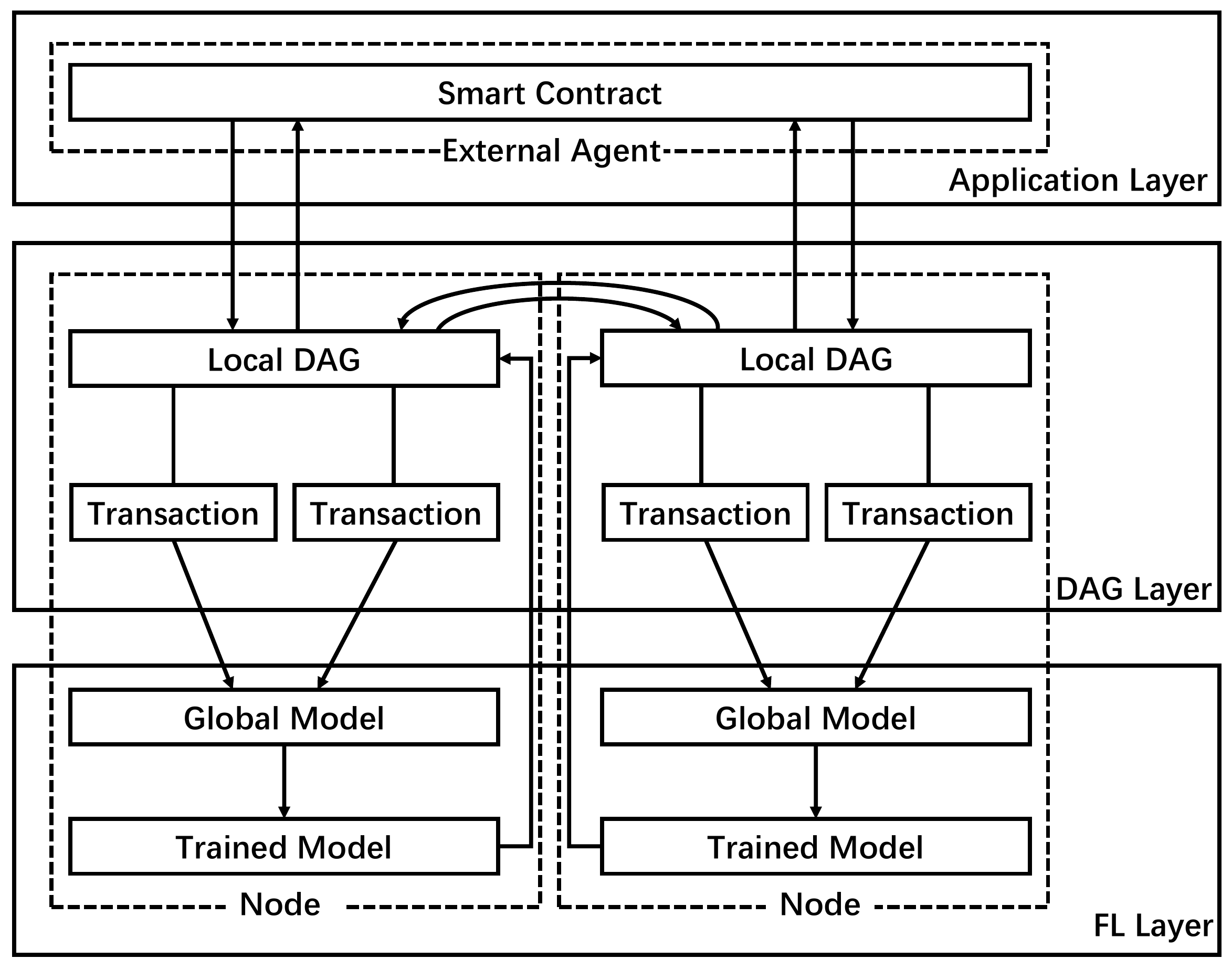}
		\caption {Architecture of DAG-FL.}
		\label{f_g}
	\end{figure}

	DAG-FL is proposed as a decentralized asynchronous FL system, including application, DAG and Fl layers from top to down. To well elaborate, this hierarchical structure of DAG-FL is shown in Fig. \ref{f_g}.
	
	\subsubsection{FL Layer}
	\par FL layer is the bottom layer to provide FL function. In order to obtain local models recorded as transactions on DAG, FL layer allows any participating node to use its own data to train the global model. The global model is formed by using \textit{FederatedAveraging} algorithm to aggregate local models stored in transactions on DAG. After training the global model, a new trained local model would be processed and published as a transaction on DAG.

	\subsubsection{DAG Layer}
	\par In DAG layer, each node maintains a local DAG, where the transaction contains authentication information, local model parameters, and the approval connections. The local DAG can be updated by broadcasting or inquiring through wireless network, and thus the new transaction (or say the new model) can be spread throughout the DAG-FL network finally.

	\subsubsection{Application Layer}
	\par Application layer is deployed on the top of DAG-FL, which provides the interface to external agents by running smart contracts. Through the smart contract, external agents can release FL task to nodes, observe FL process, and obtain the target model as soon as FL is completed. When a specific FL task is released according to a smart contract, nodes in DAG layer can participate in this work based on an incentive mechanism to gain an amount of reward \cite{salah2019blockchain}. And then, during FL process, the smart contract will observe transactions on DAG to determine whether a target model has been published.

	Accordingly, the application layer provides an interface for external agents to deploy DAG-FL easily, the DAG and FL layer form an asynchronous platform for FL. Nodes with device asynchrony like smart phones and IoT devices in DAG-FL all need to maintain a local DAG to record transactions published by every node. The local DAG on each node is updated by communicating with adjacent nodes periodically, and thus newly published transactions can be seen by all nodes. As there is no central server in DAG-FL, one node in DAG-FL constructs a global model from its local DAG to iterate FL instead of requiring a global model from the central server. This feature promises that a node in DAG-FL can immediately participate in an iteration of FL whenever it is in idle state. When the node completes an iteration of FL and gets a new trained local model, the new local model can be published on its local DAG as a transaction immediately, and latter the new published transaction would be seen by all other nodes. In this manner, the operation of any node cannot affect the state of other nodes, which would satisfy the asynchrony of mobile devices.


	%


	\subsection{Consensus based Anomaly Detection}
	\par We propose a DAG-FL consensus to keep the asynchronous FL platform stable, and present an effective way for anomaly detection in DAG-FL.
	
	\par Traditional blockchained FLs often use miners to update the blockchain by running the consensus mechanism like PoW, enabling every node to observe transactions published on blockchain. Similarly, as there are no miners in DAG-FL, each node in DAG-FL should both perform FL tasks and update the DAG by running the DAG-FL consensus. Based on the voting mechanism of DAG ledger technology, DAG-FL consensus approves the nodes by validating both the authentication and local model correctness of tips. The authentication of transactions can be validated by cryptography technology like RSA \cite{rivest1978method} in blockchain fields, and the local model can be simply validated by computing the accuracy with a test set formed by the local data. By authenticating transactions, nodes in DAG-FL can avoid transaction impersonation attacks. Furthermore, adversaries doing Sybil attacks that aim to flood the network could not impersonate other normal nodes to publish crafted transactions, which makes adversaries easy to be detected and punished. Whenever a node in DAG-FL performs one iteration of FL, it runs DAG-FL consensus and should first choose some tips on its local DAG to validate. Authenticated transactions with higher accuracy of local models would be chosen to construct the global model. The node then uses local data set to train the global model to get a local model. Finally, a new transaction that contains the newly trained local model is published and approves the tips which are used to construct the global model. Unlike the PoW consensus in other blockchained FL systems which consumes resource on solving hash cryptography problems, DAG-FL consensus avoids the extra consumption of resource unrelated to FL. 
	
	\par DAG-FL consensus combines the voting mechanism of DAG ledger technology with the process of local model validation in FL, which can effectively detect abnormal nodes and mitigate their impact on FL. With the continuous extension of transactions on DAG, every approval of a transaction means that the local model on the approved transaction is selected to form a global model and influences the target model co-construction of FL. Consequently, the more approvals a transaction get, the greater impact it will have on FL, otherwise it will be isolated and has less impact on FL. Due to this unique consensus, the machine learning model in DAG-FL is always trained towards the direction that most nodes expect, and we assume that most nodes in DAG-FL are normal nodes while only a few nodes are abnormal. Abnormal transactions published by abnormal nodes usually have less prediction accuracy on test set than that of transactions published by normal nodes. Thus, compared with normal ones, the probability of abnormal transactions being approved by subsequent published transactions is much smaller. During the process of FL, abnormal transactions are isolated and their impact is minimized. In addition, nodes with too many isolated transactions can be detected by the DAG-FL as abnormal nodes, and then DAG-FL can react to these abnormal nodes.

	\subsection{FL Algorithm}
	DAG-FL is an asynchronous FL system without any central server, so we design a special FL algorithm to perform FL iterations. This part will give some numerical definitions and introduce the FL algorithm of our DAG-FL.
	
	\par Our DAG-FL is deployed on nodes which are mobile devices under a wireless network, such as the smart phones, wearable devices, and IoT devices. We assume that these nodes can communicate with each other considering an average communication bandwidth $B$ under the wireless network. Let the set of mobile devices be denoted as \(D = \{ 1,2,3, \cdot  \cdot  \cdot ,{N_D}\} \) with \(\left| D \right| = {N_D}\). 
	\({D_i}\) is the $i$-th node in $D$, and the set of training data on $D_i$ is denoted as $S_i$ with \(\left| {{S_i}} \right| = {N_i}\), where $N_i$ is the number of samples in $S_i$. Considering different computing capacities of nodes, let $f_i$ be the processor frequency of $D_i$, used to represent the computing ability. $D_i$ should create a local DAG $g_i$ that is only visible to itself, and $g_i$ can be periodically updated. Let the model stored in the transaction be denoted as $\omega$. So the local model trained by $D_i$ at time $t$ can be denoted as $\omega^t_i$. 
	
	
	\par In order to build a common machine learning model with DAG-FL, e.g., a two-layer CNN model, usually thousands of FL iterations are requested. Initially, $D_i$ starts an FL iteration at $t_0$ by validating some tips on its local DAG first, and then choose $k$ tips with local models $\omega_{d_1}^{t_1}$, $\omega_{d_2}^{t_2}$,..., $\omega_{d_k}^{t_k}$ ($t_1,t_2,...,t_k \le t_0$, $d_1,d_2,...,d_k\in D$) to aggregate a global model $\omega^{t_0}$ using the \textit{FederatedAveraging} algorithm:
	\begin{equation}\label{fed_avg}
	\omega^{t_0} = \sum_{i=1}^{k}{n_i \omega_{d_i}^{t_i} },
	\end{equation}
	where $\sum_{i=1}^{k}{n_i}=1$, and $n_i$ is the weight factor representing the importance of local models, and to simplify the FL algorithm of our DAG-FL, here we set $n_i= 1/k$ which means each local model is equally important.
	
	After getting the global model, $D_i$ extracts $m$ samples from data set $S_i$ as a mini-batch $z_i$ to train the global model for $\beta$ epochs. Once $D_i$ gets a new local model \(\omega _i^{t_0}\) trained by \({\omega ^{t_0}}\), a transaction with $\omega _i^{t_0}$ is published on $g_i$, and the FL iteration is completed. Samples in mini-batch $z_i$ can be denoted as $(x_i,y_i)$, where $x_i$ is the feature set and $y_i$ is the label set. Then the loss function in machine learning can be denoted as \({f_{z_i}}(\omega )\),  where \({f_{z_i}}(\omega ) = l({x_i},{y_i};\omega )\), and is the prediction error of \(({x_i},{y_i})\) in $\omega$. 
	
	In addition, due to the incentive mechanism of DAG-FL, $D_i$ in DAG-FL expects to train global models in every participated FL iteration to get the local model $\omega_{i}$, which can minimize \({F_i}(\omega_{i} )\) as follows.
	\begin{equation}\label{sig_min}
	\mathop {\min }{F_i}(\omega_{i} ) = {E_{{z_i}\sim{S_i}}}{f_{{z_i}}}(\omega_i ).
	\end{equation}
	\par And for the whole DAG-FL system, external agents expect to get target model $\omega$ through smart contract after thousands of FL iterations to minimize $F(\omega )$ as follows.
	\begin{equation}\label{all_min}
	\mathop  {\min }F(\omega ) = \frac{1}{{{N_D}}}\sum\nolimits_{i \in D} {{F_i}(\omega_{i})}.
	\end{equation}
	
	\begin{figure*}[!t]
		\centering
		\includegraphics[width=7in]{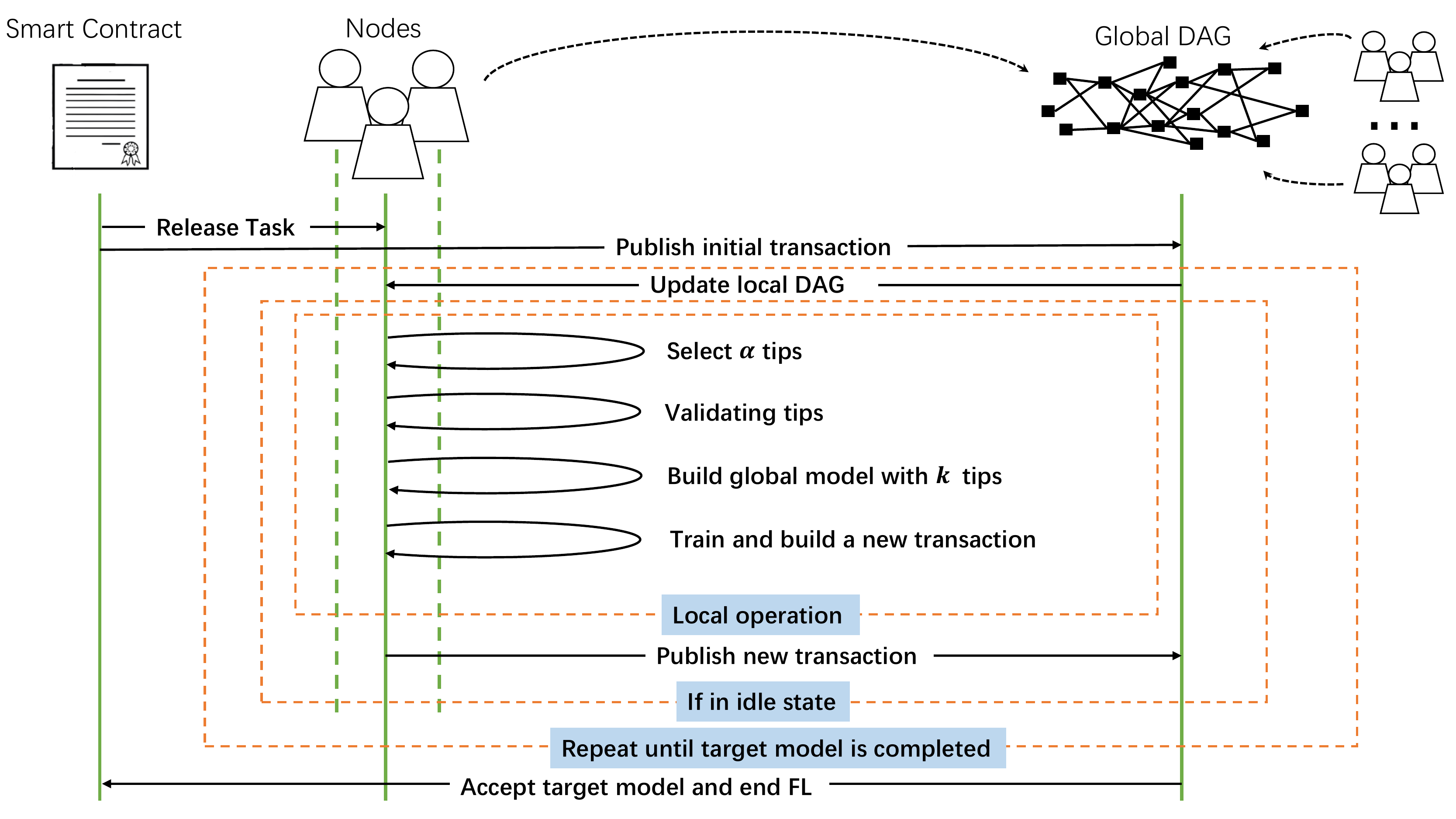}
		\caption{Sequence diagram for operations of DAG-FL.}
		\label{s_d}
	\end{figure*}

	\subsection{DAG-FL Operation}
	After a brief overview of DAG-FL, two algorithms involved in DAG-FL will be proposed in this part to introduce the operation process of DAG-FL in details.

	\par The task publisher of FL in DAG-FL can be denoted as an external agent $E$, which can be regarded as an authoritative organization in the application layer holding a virtual machine to run the smart contract. Through the smart contract, $E$ executes \textit{DAG-FL controlling} algorithm including three main functions which are (i) initializing the machine learning model to publish an initial transaction, (ii) periodically updating a local DAG to observe the FL process, (iii) and informing every node to terminate FL task after the target model is obtained. In detail, when $E$ publishes an FL task specifying the structure of the machine learning model with an expected final prediction accuracy $AC{C_0}$, nodes in $D$ are applied to participate in this FL task. At beginning time of the FL task $t_0$, $E$ publishes the initial transaction with ${\omega _0^{t_0}}$. After $t_0$, $E$ updates its local DAG $g_0$ and randomly selects $\alpha$ tips within ${\tau _{\max }}$ to validate. After validation, $k$ tips with the highest accuracy will be selected to construct a global model ${\omega_0}$. If prediction accuracy of the $\omega_0$ $ACC_t$ $\geq$ $ACC_0$, $E$ broadcasts the end signal to all participating nodes in DAG-FL to terminate the FL, and ${\omega_0}$ is the final target model. Otherwise, $E$ will repeat to update $g_0$ until the proper $\omega_0$ is found.
	
	\begin{algorithm}[h]
		\caption{\textbf{DAG-FL Controlling.} $ACC_0$ is the expected prediction accuracy of target model, $D$ is the set of nodes, $g_0$ is the local DAG on $E$, ${\tau _{\max }}$ is the threshold staleness of tips.}
		\label{alg:A}
		\text{\textbf{External agent $E$ executes:}}
		\begin{algorithmic}[1]
			\STATE {\textbf{Input} $ACC_0$}
			\STATE {$\omega_0^{t_0}$ $\gets$ Initialize the machine learning model  }
			\STATE {Publish initial transaction including $\omega_0^{t_0}$ to $D$}
			\WHILE {true}
			\STATE {Update $g_0$}
			\STATE {Validate $\alpha$ tips on local DAG $g_0$ within ${\tau _{\max }}$}
			\STATE {$\omega_0$ $\gets$ Select $k$ tips with the highest accuracy to compute global model according to Eq. (\ref{fed_avg})}
			\STATE {$ACC_t$ $\gets$ Get accuracy by $\omega_0$}
			\IF {$ACC_t>ACC_0$}
			\STATE {Send end signal to $D$}
			\STATE {\textbf{break}}
			\ENDIF
			\ENDWHILE
			\STATE {\textbf{Output} $\omega_0$}
		\end{algorithmic}
	\end{algorithm}
	\begin{algorithm}[h]
		\caption{\textbf{DAG-FL Updating.} $D$ is the set of nodes, $g_i$ is the local DAG on $D_i$, ${\tau _{\max }}$ is the threshold staleness of tips, $S_i$ is the local data set of $D_i$.}
		\label{alg:B}
		\text{\textbf{Node $D_i$ ($D_i \in D$) executes:}}
		\begin{algorithmic}[1]
			\WHILE {true}
			\IF {end signal is recieved}
			\STATE {\textbf{break}}
			\ELSE 
			\STATE {Update $g_i$}
			\IF {idle state}
			\STATE {Validate $\alpha$ tips on local DAG $g_i$ within ${\tau _{\max }}$}
			\STATE {$\omega_i$ $\gets$ Choose $k$ tips with the highest accuracy to compute global model according to Eq. (\ref{fed_avg})}
			\STATE {$\omega_i^{t}$ $\gets$ Train $\omega_i$ with $S_i$ for $\beta$ epochs}
			\STATE {Publish the new transaction including $\omega_i^{t}$ and approvals on $g_i$}
			\ENDIF
			\ENDIF
			\ENDWHILE
		\end{algorithmic}
	\end{algorithm}

	\par After $t_0$, all the participating nodes run the \textit{DAG-FL updating} algorithm to achieve DAG-FL consensus whenever they are in idle state. As nodes in DAG-FL are mobile devices, nodes may shut down during one iteration of FL because of device asynchrony. And any node in DAG-FL which is in idle state should go through four stages to participate within one iteration:  
	\subsubsection{Stage 1}
	The node selects some tips (no more than $\alpha$) within an appropriate staleness ${\tau _{\max }}$ from local DAG randomly. 
	\subsubsection{Stage 2}
	The node first validates the authentication of tips selected in stage 1. Then the node computes the prediction accuracy of the local models in the selected tips using its own test data set.  
	\subsubsection{Stage 3}
	The node chooses $k$ ($k<\alpha$) tips selected in the first stage with the highest accuracy to run \textit{FederatedAveraging} algorithm and gets a global model. The node utilizes local data set to train the global model and gets a trained local model.
	\subsubsection{Stage 4}
	A new transaction is constructed. The new transaction contains authentication information, local model trained in the third stage, and the approval information to approve the $k$ tips chosen in the third stage.
	\par After the completion of the above four stages, the node then successfully finishes one iteration in DAG-FL and publishes the new transaction to the DAG, which will be soon observed by all other nodes in DAG-FL. For clarity, we elaborate the sequence diagram for operations of DAG-FL in Fig. \ref{s_d}. Global DAG is a collection of local DAGs and is virtual. The process of local DAG updating can be seen as communicating with the global DAG.

	\par
\begin{figure}[!t]
		\centering
		\includegraphics[width=3in]{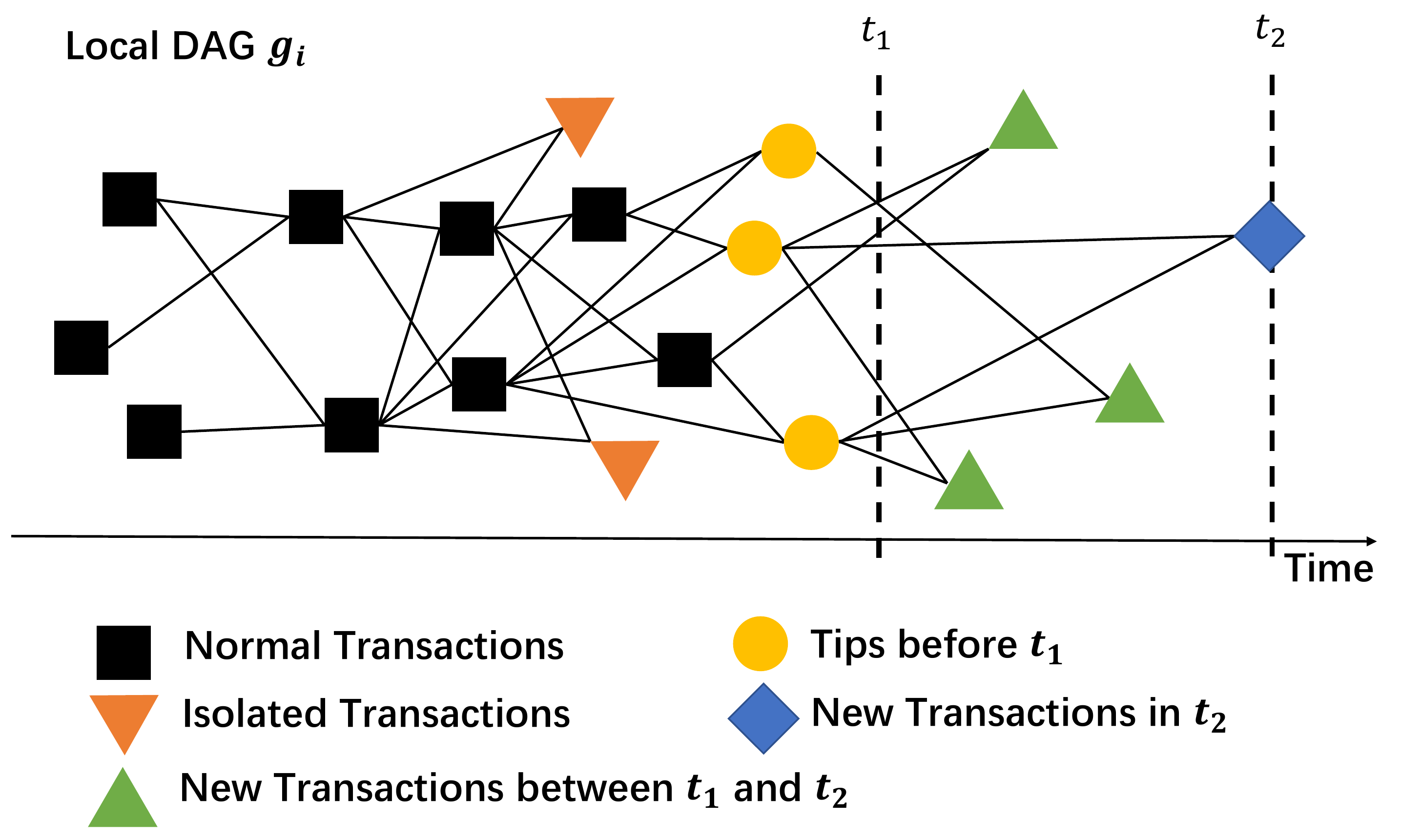}
		\caption{Transaction alternations on DAG $g_i$.}
		\label{dag_alt}
	\end{figure}

 Here we take node $D_i$ in DAG-FL as an example to illustrate the alterations of transactions on $g_i$ during the process of running the \textit{DAG-FL updating}. Suppose that $D_i$ is in idle state at $t_1$ and intends to perform an iteration of DAG-FL. As shown in Fig. \ref{dag_alt}, $D_i$ selects $\alpha$ recently published and has not been yet approved tips from its local DAG $g_i$ to validate them. $D_i$ then computes the global model $\omega_i$ with $k$ validated transactions by using the \textit{FederatedAveraging} algorithm. After $D_i$ trains $\omega_i$ with local data set $S_i$ for $\beta$ epochs, a new local model $\omega_i^{t_2}$ is built, at time $t_2$. Finally, a new transaction contained $\omega_i^{t_2}$ and approval information is published to $g_i$.

	\par As DAG-FL is a completely decentralized asynchronous FL system without any central server, a global model can only be temporarily constructed from the local DAG. Thus, DAG-FL allows nodes with different global models to train at the same time during the FL process. Global models differ a lot from each other in the early stage of FL, which causes a low convergence rate of the target model. However, these temporarily constructed global models tend to be almost the same in the later stage of FL, which promises an impressive convergence rate just as classic asynchronous FL.
	
	
	

	\section{Deployment and Stability Analysis}
	After introducing the framework and operation process of DAG-FL, in this section, we present two crucial factors that directly affect the stability of the system when deploying DAG-FL.
	\subsection{Tips on DAG}
	In DAG-FL, global models are temporarily constructed by tips on DAG. If the number of tips on a local DAG is too large simultaneously, global models constructed by tips can be significantly different from each other. This will significantly reduce the efficiency of DAG-FL in the early stage, and even the co-constructed machine learning model may never converge. For different FL tasks, maintaining the number of tips around a constant value $L_0$ ($L_0>0$) at any time is the key to ensure DAG-FL to stably and efficiently operate.
	\par We assume that the average probability of nodes in DAG-FL successfully completing an iteration under the wireless network is $p$. Meanwhile, the number $n$ of nodes in DAG-FL is quite large, and the process of successfully participating an iteration for nodes can be regarded as a Poisson process with the arrival rate $\lambda$. Therefore, the arrival rate can be modeled as $\lambda=np$, which indicates that $\lambda$ nodes start to do one iteration of DAG-FL per unit time. In order to facilitate the analysis, we assume $\lambda$ is known in advance during the process of FL, and the average time for a node to complete an iteration is $h$ in DAG-FL. According to the derivation in tangle \cite{popov2016tangle}, we get:
	\begin{equation}\label{p_p}
	{L_0} = \frac{{k\lambda h}}{{k - 1}}.
	\end{equation}
	
	\par In an FL task, the training file size of a mini-batch is denoted as ${\varphi _{0}}$, and the file size of the data set for validating other transactions is ${\varphi _{1}}$. The computation density of the training global model is denoted as ${\eta _{0}}$, and the computation density of validating a transaction is denoted as ${\eta _{1}}$. 
	Based on the above definitions, the training delay $d_0$ of an iteration for node $D_i$ can be computed as
	\begin{equation}\label{c_d}
	{d_0} =  \frac{{{\eta _{0}} \times {\varphi _{0}} \times \beta }}{f}.
	\end{equation}
	
	\par Next, the delay $d_1$ caused by validating can be approximately expressed as
	\begin{equation}\label{cc_d}
	{d_1} = \frac{{{\eta _{1}} \times {\varphi _{1}} \times \alpha }}{f}.
	\end{equation}
	\par In summary, the delay of the whole iteration $h$ can be approximately expressed as
	\begin{equation}\label{h_d}
	{h} = d_0+d_1. 
	\end{equation}
	
	\par Substituting Eq. (\ref{h_d}) into  Eq. (\ref{p_p}), we can get $L_0$ as follows.
	\begin{equation}\label{L}
	{L_0} = \frac{{k\lambda ({\eta _{0}} \times {\varphi _{0}} \times \beta  + {\eta _{1}} \times {\varphi _{1}} \times \alpha )}}{{(k - 1)f}}.
	\end{equation}
	
	\par For a specific FL task, $h$ and $\lambda$ can be regarded as constant values. Thus, according to Eq. (\ref{L}), $L_0$ is mainly determined by $k$ and $\alpha$. If tips in DAG-FL are too large, we can set a larger $k$ to reduce $L_0$ and keep DAG-FL running stably. It is worth noting that $k<\alpha$, and the setting of $k$ and $\alpha$ is also related to the immunity of abnormal nodes in DAG-FL. For fixed $\alpha$, we can observe that the larger the $k$ will lead to the less probability that a transaction published by an abnormal node is isolated. Thus, when deploying DAG-FL to run a specific FL task, we can weigh and balance $k$ and $\alpha$ to keep DAG-FL operating with a proper $L_0$.
	
	\subsection{Staleness of Transactions}
	The interval between the publishing time of a transaction and the current time is called staleness \cite{xie2019asynchronous}, denoted as $\tau $. Obviously, the longer interval causes the worse the staleness. Generally, the transactions with better staleness can build a global model with better progress of FL.

	\par With the assistant of the global model of FL, we can train a local model that is closer to the target model. If a node uses local data set to train a global model constructed by tips with bad staleness and gets a local model, the new transaction which contains the trained local model is more likely to be isolated on DAG. These isolated transactions published by normal nodes are not conducive to the machine learning model convergence, and significantly slow down the establishment of the target model of FL. In order to get a global model that can well represent the progress of FL, we need to select tips with good staleness on DAG to construct the global model in an iteration of DAG-FL.

	\par In order to prevent nodes in DAG-FL from using transactions with bad staleness when constructing the global model, we set a threshold ${\tau _{\max }}$ of staleness for the tips on DAG. Consequently, nodes will select tips with staleness no more than ${\tau _{\max }}$ to construct the global model for training. And once the staleness of a transaction exceeds ${\tau _{\max }}$, the transaction cannot be used as a tip on local DAG anymore. It should be noted that though a small ${\tau _{\max }}$ can ensure that tips all have good staleness, the number of tips $L_0$ may sharply reduce, which is also not conducive to the normal operation of DAG-FL. Thus, we need to weigh and balance the ${\tau _{\max }}$ when deploying DAG-FL to solve a specific FL task.
	
	\section{Simulation and Implementation}
	In order to evaluate the performance of DAG-FL, in this section, we first design a simulation platform to verify the performance of DAG-FL with large-scale participating nodes, and then conduct the actual deployment of DAG-FL on a small number of nodes using the proposed algorithm.
	
	\subsection{Simulation}
	\subsubsection{Platform Setting}
	We design a simulation platform pySimuFL that can evaluate the performance of the proposed DAG-FL as compared to Google FL \cite{mcmahan2017communication}, Asynchronous FL \cite{xie2019asynchronous} and Block FL \cite{kim2019blockchained}. For simulation, there are 100 nodes in our pySimuFL that share the same wireless network with a radius of 10km. In order to illustrate the performance of DAG-FL more comprehensively, we consider a CNN task for image classification and an LSTM task for language modeling in the following simulation experiments.

	\begin{figure*}[tbp]
		\centering
		\subfloat[]{\includegraphics[width=2.4in]{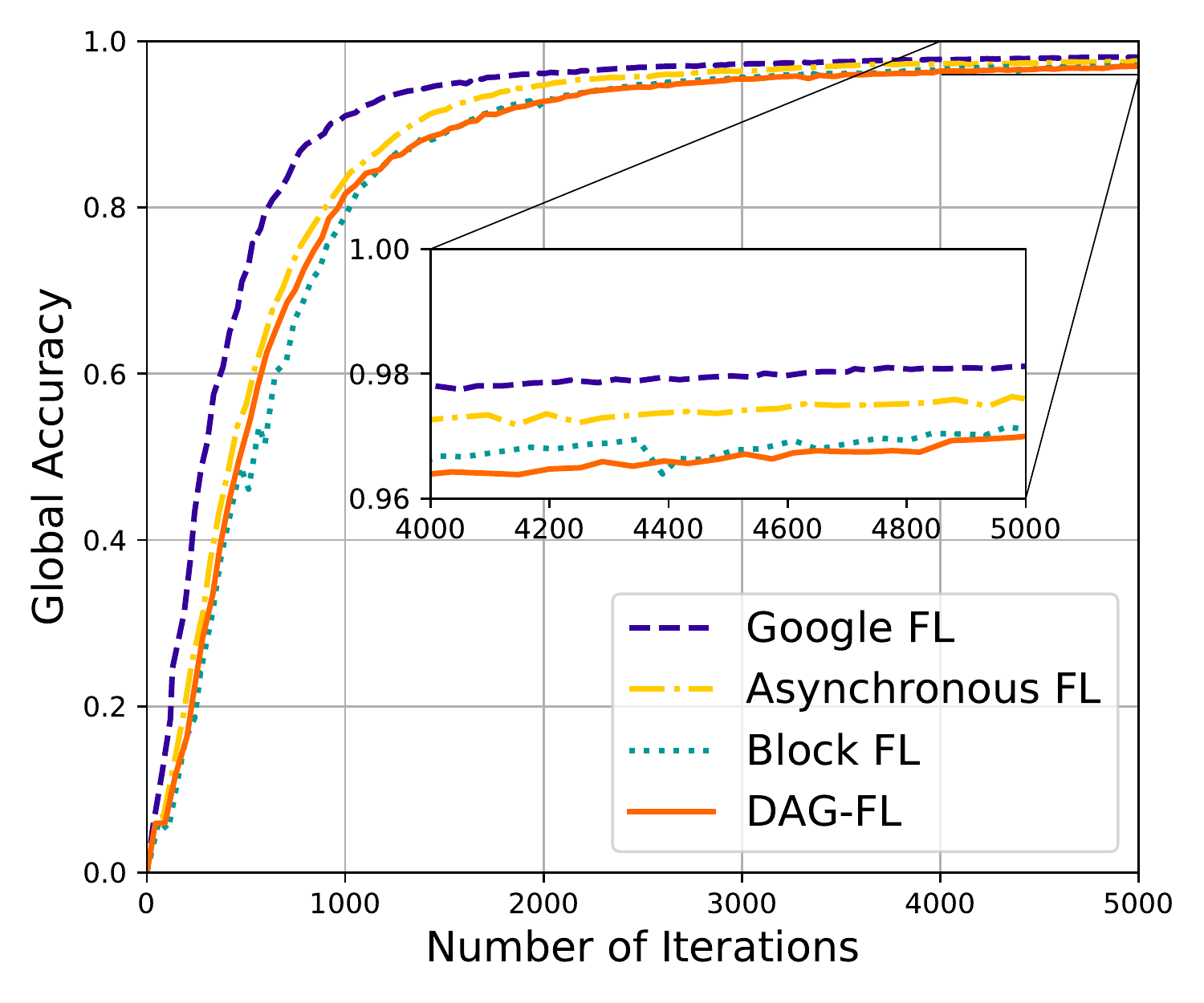}\label{cnn_normal_acc}} 
		\subfloat[]{\includegraphics[width=1.2in]{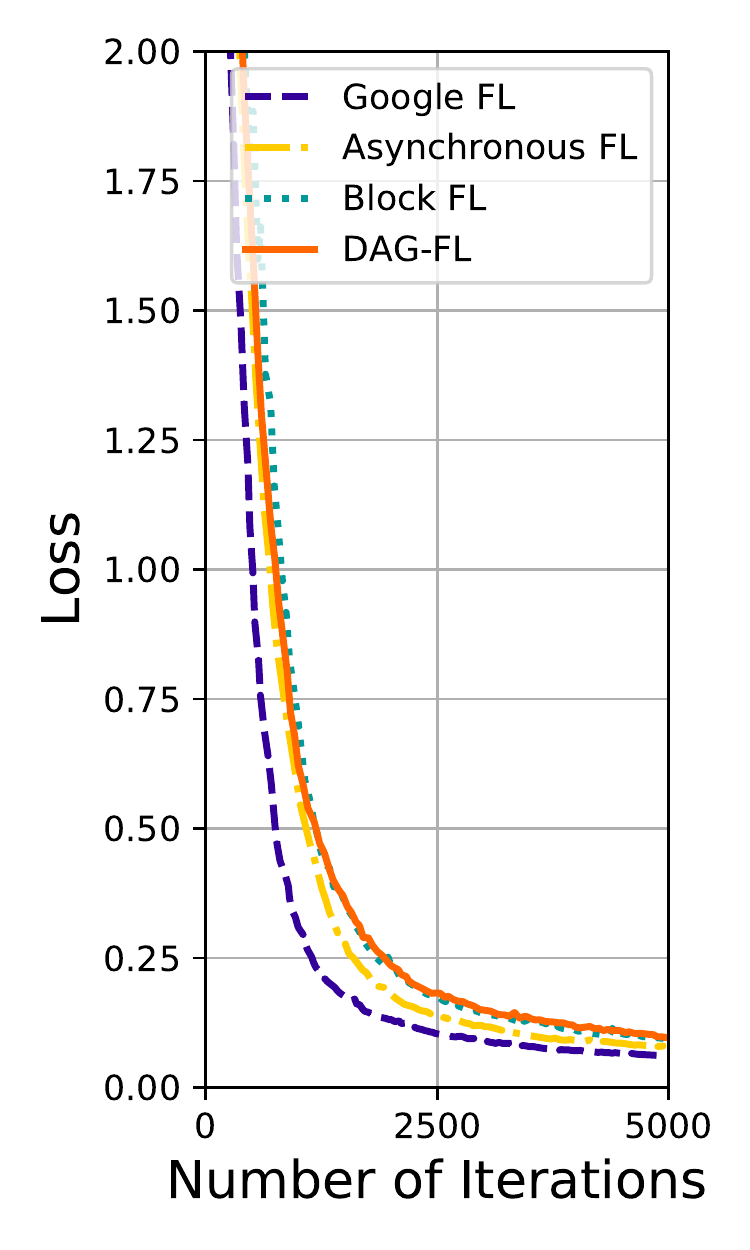}\label{cnn_normal_loss}}
		\subfloat[]{\includegraphics[width=2.4in]{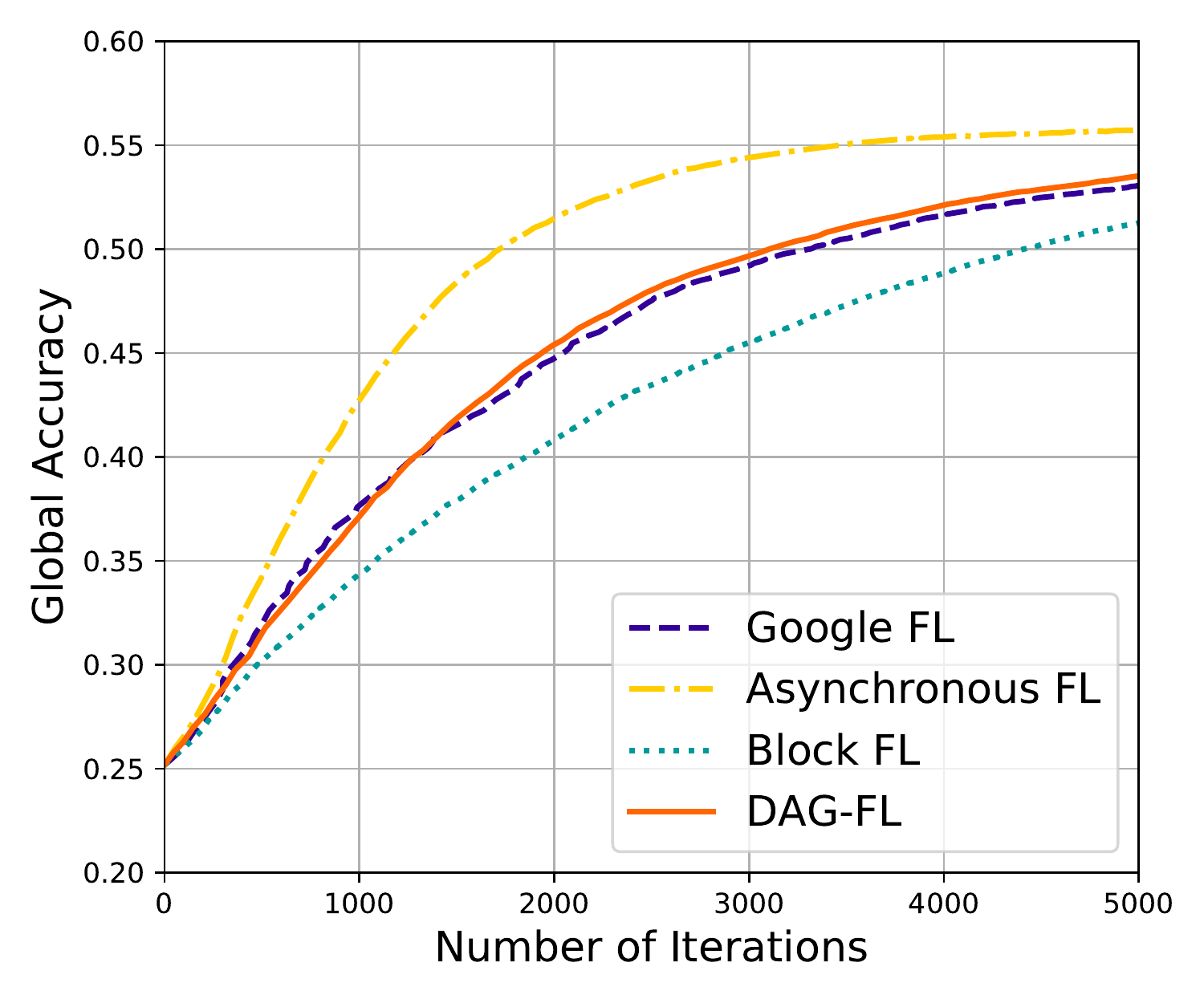}\label{lstm_normal_acc}}
		\subfloat[]{\includegraphics[width=1.2in]{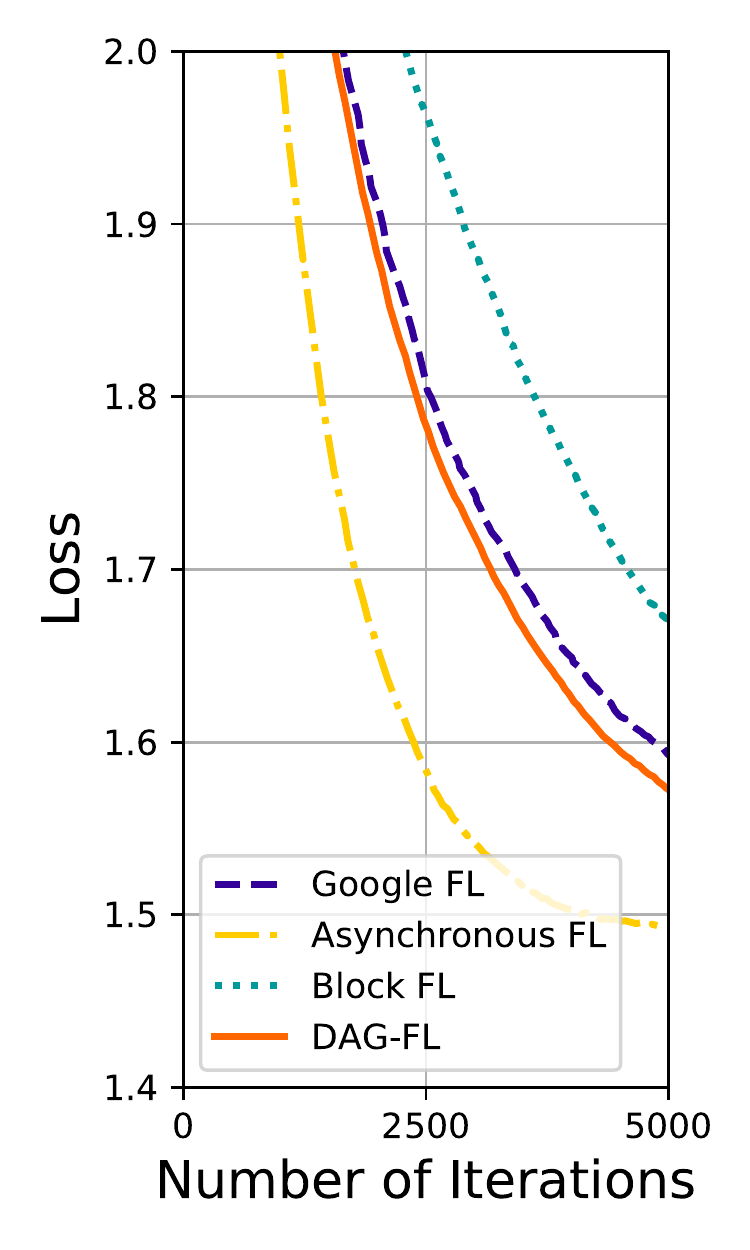}\label{lstm_normal_loss}}
		\caption{(a): The test accuracy of different federated learning systems in ideal case for CNN task; (b): The training loss of different federated learning systems in ideal case for CNN task; (c): The test accuracy of different federated learning systems in ideal case for LSTM task; (d): The training loss of different federated learning systems in ideal case for LSTM task.}
		\label{exp_normal}
	\end{figure*}

	For CNN task, a CNN model proposed in \cite{mcmahan2017communication} which has two 5x5 convolution layers (the first with 32 channels, the second with 64, each followed with 2x2 max pooling), a fully connected layer with 512 units ReLu activation, and a final softmax output layer is conducted using the MNIST dataset. The MNIST dataset of handwritten digits has a training set of 60,000 samples and a test set of 10,000 samples. To reflect the non-IID feature of local data on mobile devices, we design a special scheme to distribute training data to every node. At first, we sort 2/3 training set by digit label, divide it into 200 shards of size 200, and assign each of 100 nodes 2 shards. Then, the remained training set is equally assigned to 100 nodes. By using this data distributing scheme, each node is assigned a local data set including most samples of two exact digits and some samples of the other digits. In addition, the learning rate of this CNN model is set to be 0.002 and the loss function is cross-entropy.
	
	For LSTM task, a stacked character-level LSTM language model \cite{mcmahan2017communication} is conducted using the Shakespeare dataset to predict the next character of a speaking line. The LSTM model takes a series of characters with a length of 80 as input and embeds each of these into a learned 8-dimensional space. The embedded characters are then processed through 2 LSTM layers, each of which has 256 nodes. Finally, a softmax output layer is connected to the second LSTM layer with one node per character. The Shakespeare dataset is split into a train set with 3564579 characters and a test set with 870014 characters. As the Shakespeare dataset itself is composed of 1146 roles' speakings, which is already highly unbalanced, the non-IID feature of local data is realized by randomly assigning the training set to 100 nodes. The LSTM model uses the cross-entropy loss function and is set with a learning rate of 0.3. To better compare the convergence difference among FLs in the following simulation experiments, the LSTM model has been pre-trained to achieve an accuracy of 0.2518.

	Considering a quasi-static network environment, we set up a relatively conservative bandwidth of wireless network, i.e.,100 Mbps. Nodes are set to be in idle state for FL at different times, thus enabling one node on average ready for an FL iteration per second. Other parameter settings of the simulation platform are listed in Table I.
	
	
	
	\begin{figure*}[tbp]
		\centering
		\subfloat[]{\includegraphics[width=2.4in]{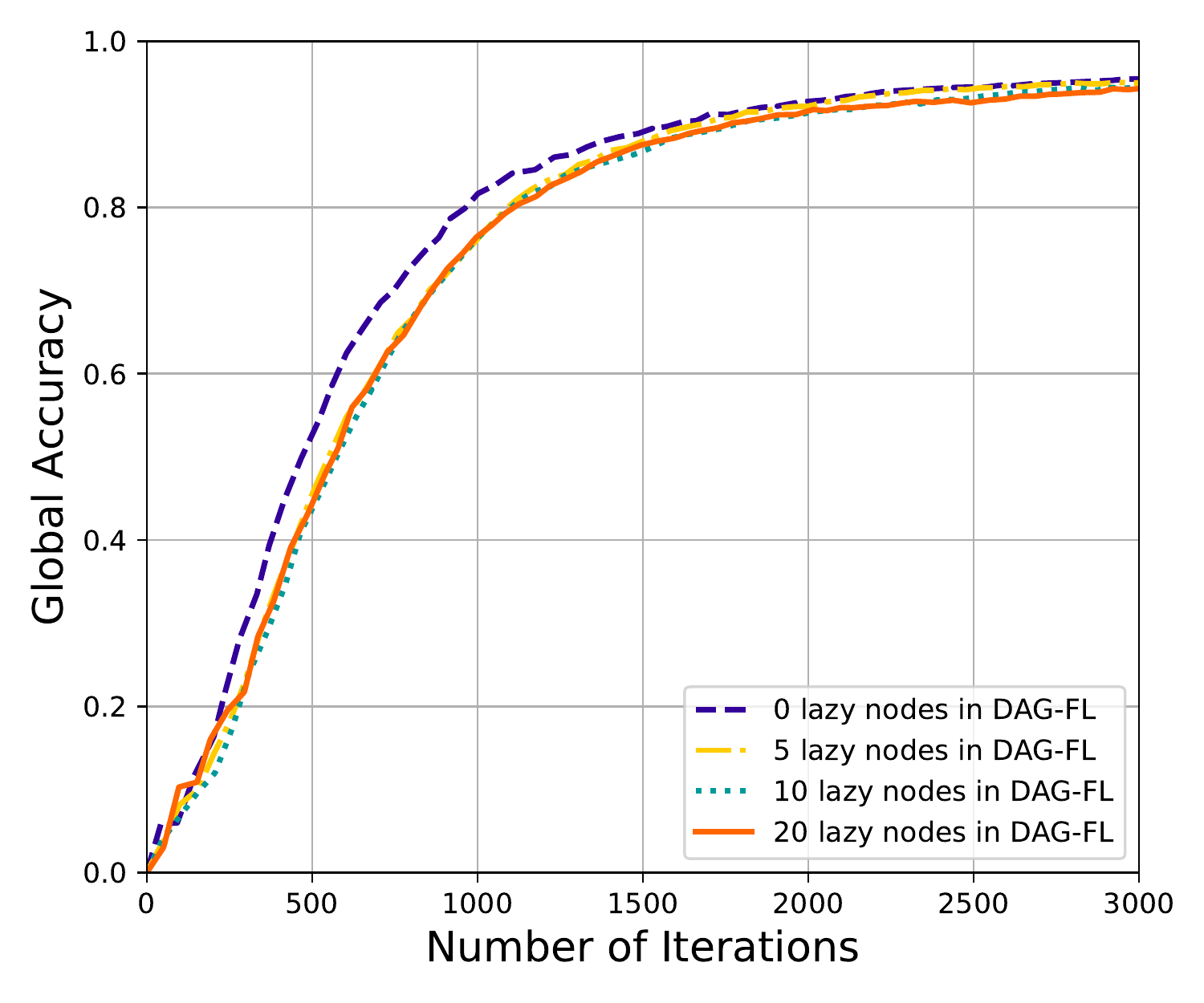}\label{cnn_com_l}} 
		\subfloat[]{\includegraphics[width=2.4in]{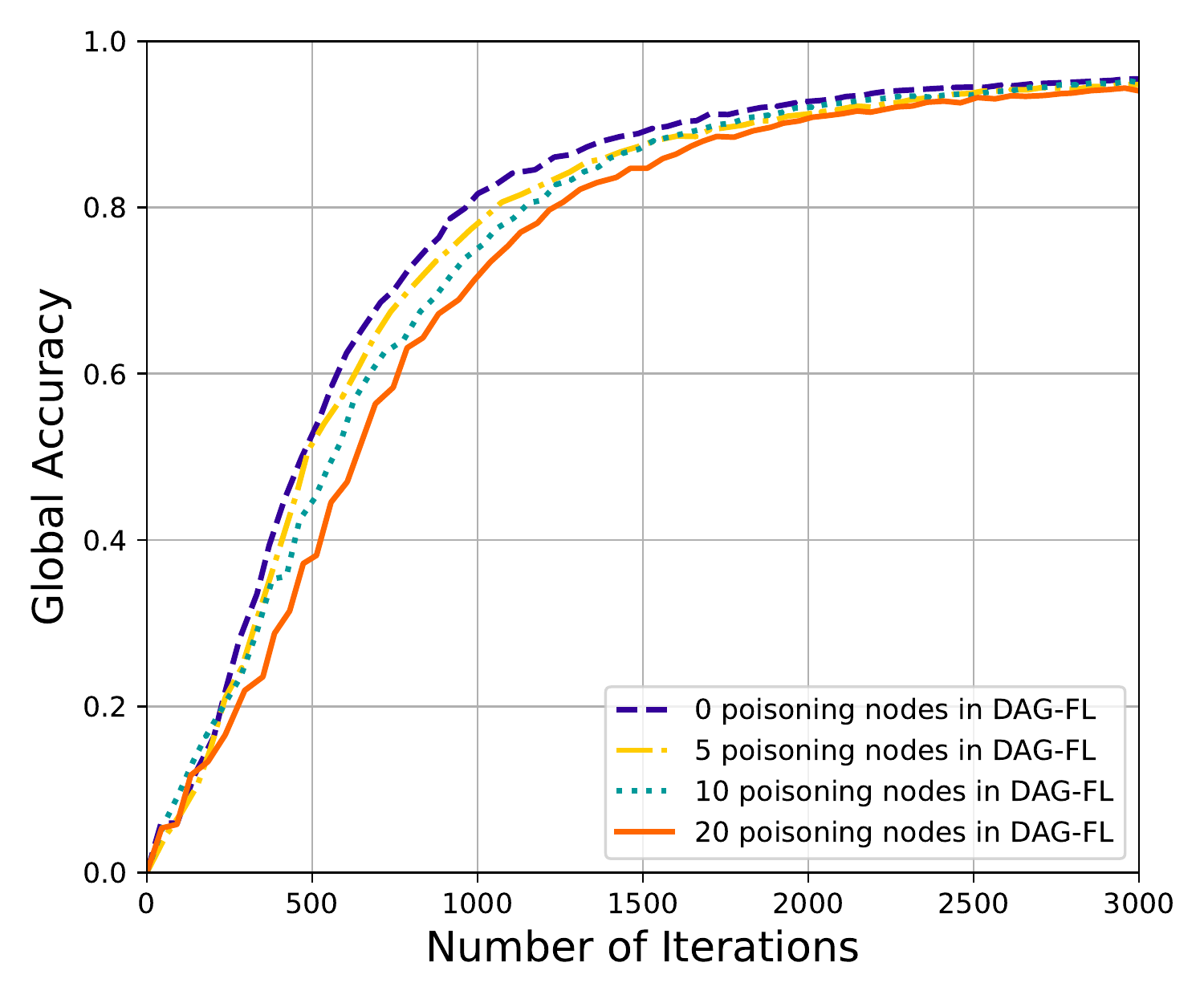}\label{cnn_com_i}}
		\subfloat[]{\includegraphics[width=2.4in]{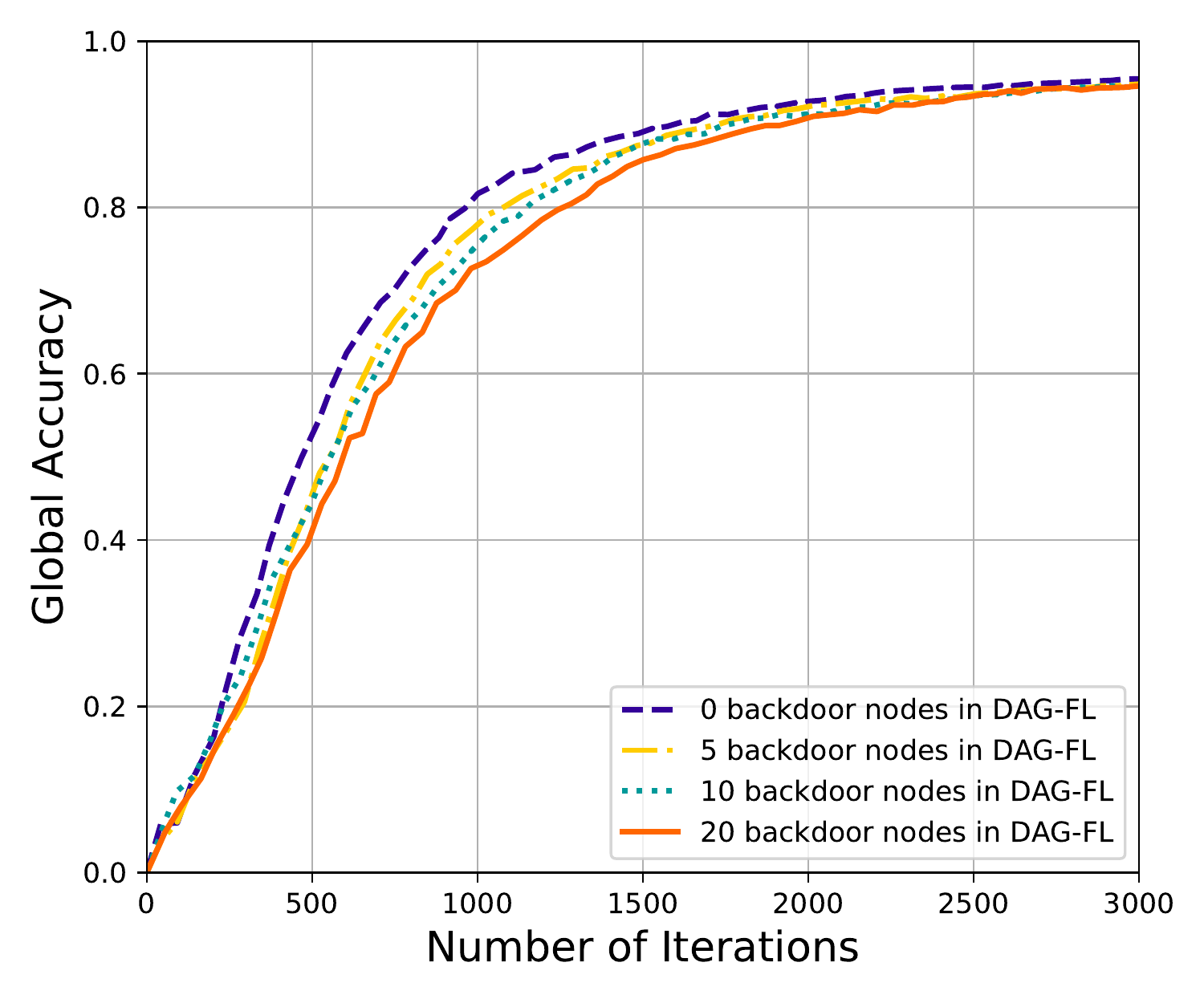}\label{cnn_com_t}}
		\caption{(a): The accuracy of DAG-FL with lazy nodes for CNN task; (b): The accuracy of DAG-FL with poisoning nodes for CNN task; (c): The accuracy of DAG-FL with backdoor nodes for CNN task.}
		\label{dag_comp}
	\end{figure*}

	\begin{table}[]
		\label{platform}
		\centering
		\caption{Platform settings}
		\begin{tabular}{|c|l|c|c|}
			\hline
			Symbol           & \multicolumn{1}{c|}{Definition}                           & CNN              & LSTM            \\ \hline
			$\phi $          & The file size of a transaction       & 7MB              & 3MB             \\ \hline
			${\varphi _{0}}$ & The file size of each minibach       & 0.3MB            & 9KB             \\ \hline
			${\varphi _{1}}$ & The file size of validation data set & 0.3MB            & 9KB             \\ \hline
			$\beta$          & Training epochs of an iteration      & 1                & 5               \\ \hline
			$m$              & Minibatch size                       & \multicolumn{2}{c|}{100}           \\ \hline
			${\eta _{0}}$    & The density of training              & \multicolumn{2}{c|}{500cycles/bit} \\ \hline
			${\eta _{1}}$    & The density of validating            & \multicolumn{2}{c|}{160cycles/bit} \\ \hline
			$f$              & CPU frequency for each node          & \multicolumn{2}{c|}{1-2GHz}        \\ \hline
			$k$              & Approved transactions                & \multicolumn{2}{c|}{2}             \\ \hline
			$\alpha$         & Chosen transactions                  & \multicolumn{2}{c|}{5}             \\ \hline
			$B$              & Bandwidth
			& \multicolumn{2}{c|}{100Mbps}       \\ \hline
			$\tau_{\max}$    & Staleness threshold                  & \multicolumn{2}{c|}{20s}           \\ \hline
		\end{tabular}
	\end{table}

	\begin{table}[]	
		\label{delay}
		\centering
		\caption{Iteration delay}
		\begin{tabular}{|c|c|c|}
			\hline
			\multicolumn{1}{|c|}{\multirow{2}{*}{FL Systems}} & \multicolumn{2}{c|}{Average latency for 100 iterations} \\ \cline{2-3} 
			\multicolumn{1}{|c|}{}                            & \multicolumn{1}{c|}{CNN}   & \multicolumn{1}{c|}{LSTM}  \\ \hline
			Google FL                                         & 150.04s                    & 144.07s                          \\ \hline
			Asynchronous FL                                   & 105.88s                    & 101.40s                          \\ \hline
			Block FL                                          & 113.91s                    & 115.49s                          \\ \hline
			DAG-FL                                            & 107.43s                    & 100.26s                          \\ \hline
		\end{tabular}
	\end{table}

	\par We consider normal and abnormal nodes in pySimuFL, where abnormal nodes can be set as lazy nodes, poisoning nodes, and backdoor nodes. Lazy nodes upload existing models instead of models trained with local data in an FL iteration, aiming to obtain potential rewards of FL. Poisoning nodes are set to have wrong data for training, which can poison the global model in FL and aim at reducing the overall target model performance \cite{zhang2019poisoning}. Backdoor nodes are those that do targeted attacks (also called backdoor attacks) and aim to mislead the target model in FL. As the LSTM model using Shakespeare dataset is vulnerable to targeted attacks, backdoor nodes are only concerned in CNN task. For the CNN task in our pySimuFL, backdoor nodes modify part of their local MNIST pictures by replacing a 5x5 white square into the upper left corner \cite{8835365} and want to mislead the final target model of FL to recognize pictures with a white square to the wrong digit label which is the true digit plus one.

	\par The results of Google FL, Asynchronous FL, and Block FL in the following figures are obtained by using the same nodes settings as DAG-FL. Note that here one iteration refers to the process that one node uses its local data set to train the global model for $\beta$ epochs. To compare the performance differences of the several FL systems fairly, some settings of Google FL, Asynchronous FL, and Block FL are presented as follows. For Google FL, in each round, ten nodes in idle state are selected to download and train the global model from the central server, which means Google FL will run ten iterations in a round. Asynchronous FL used in pySimuFL updates the global model by averaging the last global model with newly uploaded local model. And for iteration process of Block FL, we set up 100 nodes and 5 extra miners. The 100 nodes in Block FL are divided into 5 groups and each group is associated with a miner. Every miner has the whole test data set (10000 picture samples for CNN task and 10740 80-length character line samples for LSTM task) to validate published transactions. Whenever a miner collects 5 transactions published by nodes or waits over 10s, it runs PoW consensus mechanism to win the right to publish the next block (where the global model stores) into blockchain. We ignore the fork problem of Block FL, and set up a small difficulty for PoW which in average costs miner 5s to solve a hash cryptography problem. The average time consumption per 100 iterations for the four FLs is shown in Table II.

	\subsubsection{Analysis of DAG-FL efficiency}
	We first evaluate the test accuracy and training loss of DAG-FL compared with other FL systems in the ideal case. 
	
	\par Figure \ref{exp_normal} displays the test accuracy and training loss of the four FLs on both CNN and LSTM tasks. For CNN task, DAG-FL and Block FL have nearly the same convergence rates and have lower accuracy than Google FL and Asynchronous FL at the first 5000 iterations. This is because that Google FL and Asynchronous FL can promise an equal frequency for data on each node to be used. For LSTM task, Asynchronous FL and DAG-FL have better convergence rates than the synchronous Google FL and pseudo-asynchronous Block FL at the first 5000 iterations, and Asynchronous FL performs better than DAG-FL because it ensures the average use of data. Considering the iteration latency of the four FLs in Table. II, DAG-FL and asynchronous FL can complete much more FL iterations than Google FL and Block FL after the same time. When time comes to 10000s, in CNN task, all the four FLs can train a target model that achieves the accuracy around 0.982. And in LSTM task, DAG-FL and Asynchronous FL can get target models with an accuracy of 0.552 while the other two FLs can only get final target models with the accuracy below 0.548. Through comprehensive consideration of the convergence rate and average iteration latency, we conclude that DAG-FL is an efficient FL system and can train a final target model with good accuracy. 
	

	\subsubsection{Discussion on the immunity of abnormal nodes influence}
	By setting different numbers of nodes as lazy nodes, poisoning nodes, and backdoor nodes, we get accuracy curves of DAG-FL with abnormal nodes. For CNN task, the accuracy of different numbers of abnormal nodes in these three cases is shown as Fig. \ref{dag_comp}.
	
	\par In DAG-FL, the transactions published by abnormal nodes are more likely to be isolated and have less influence on the target model co-construction. The results in Fig. \ref{dag_comp} confirm this inference, in which DAG-FL is insensitive to the impact of abnormal nodes including lazy, poisoning, and backdoor ones. Even 20 percent of nodes are set to be abnormal, the convergence rate of DAG-FL just slightly reduces at the early stage, and DAG-FL can co-construct a target model of good accuracy at 3000 iterations.  
	
	\par
\begin{figure}[!t]
		\centering
		\includegraphics[width=3in]{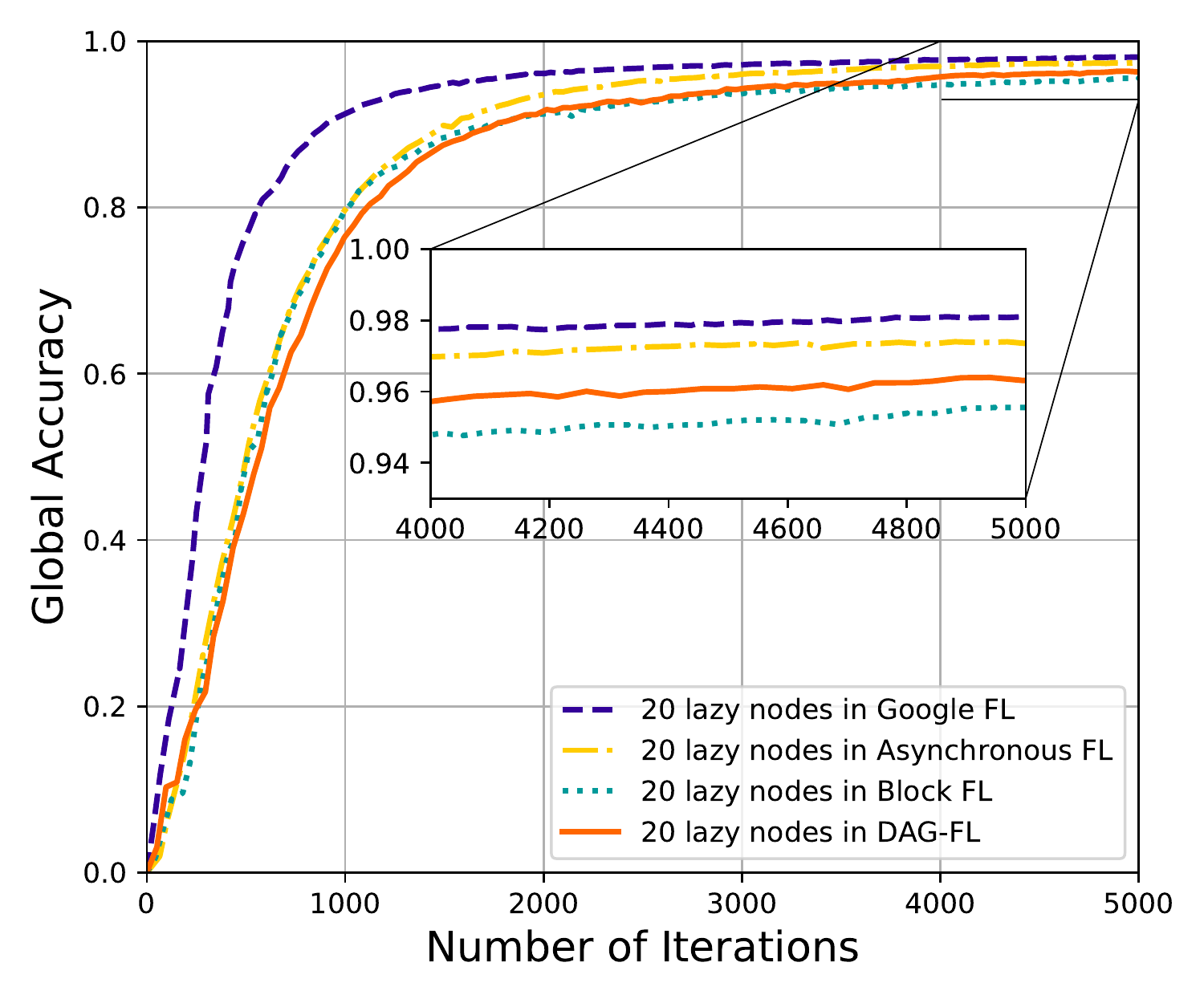}
		\caption{The accuracy of different FL systems with lazy nodes for CNN task.}
		\label{cnn_l}
	\end{figure}

\begin{figure}[!t]
		\centering
		\includegraphics[width=3in]{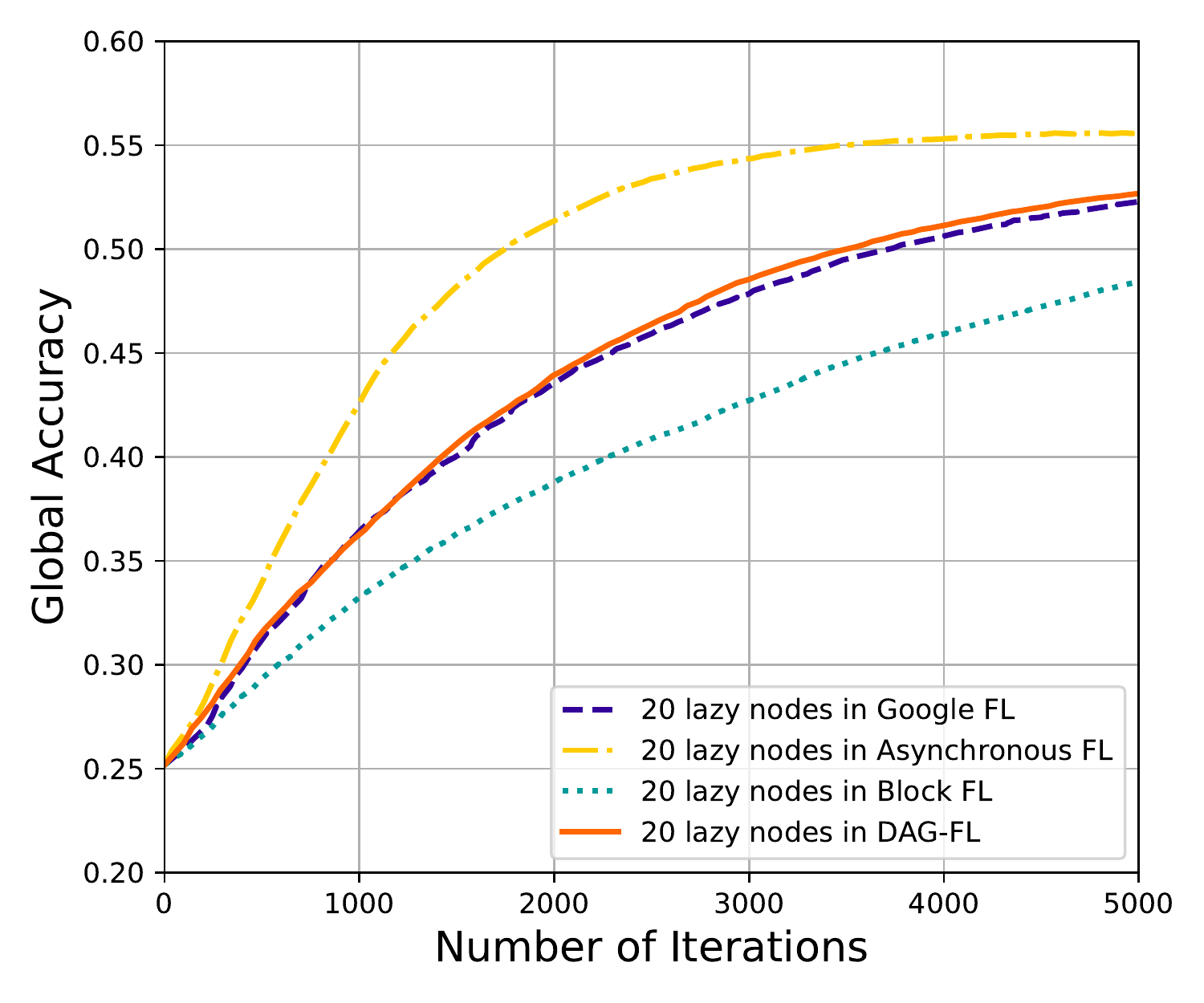}
		\caption{The accuracy of different FL systems with lazy nodes for LSTM task.}
		\label{lstm_l}
	\end{figure}

 The accuracy of different FLs with 20 percent lazy nodes for CNN and LSTM tasks is shown in Fig. \ref{cnn_l} and Fig. \ref{lstm_l}. We find that a small number of lazy nodes in Google FL and Asynchronous FL can help the global model avoid sharp updating during the FL process. Thus, these two FLs still have good convergence rates when 20 percent lazy nodes are involved for both CNN and LSTM tasks. It can be observed that Block FL is significantly affected by the participation of lazy nodes in terms of the convergence rate and the accuracy. Compared with the accuracy at 5000 iterations in the ideal case, Block FL loses about 0.02 accuracy in CNN task and about 0.04 accuracy in LSTM task when 20 lazy nodes are involved. The reason is that with more associated lazy nodes, the miner in Block FL usually has more time to run PoW to win the right to publish the next block. This means that more transactions published by normal nodes are dropped instead of being adopted for the global model updating.

	\par
\begin{figure}[!t]
		\centering
		\includegraphics[width=3in]{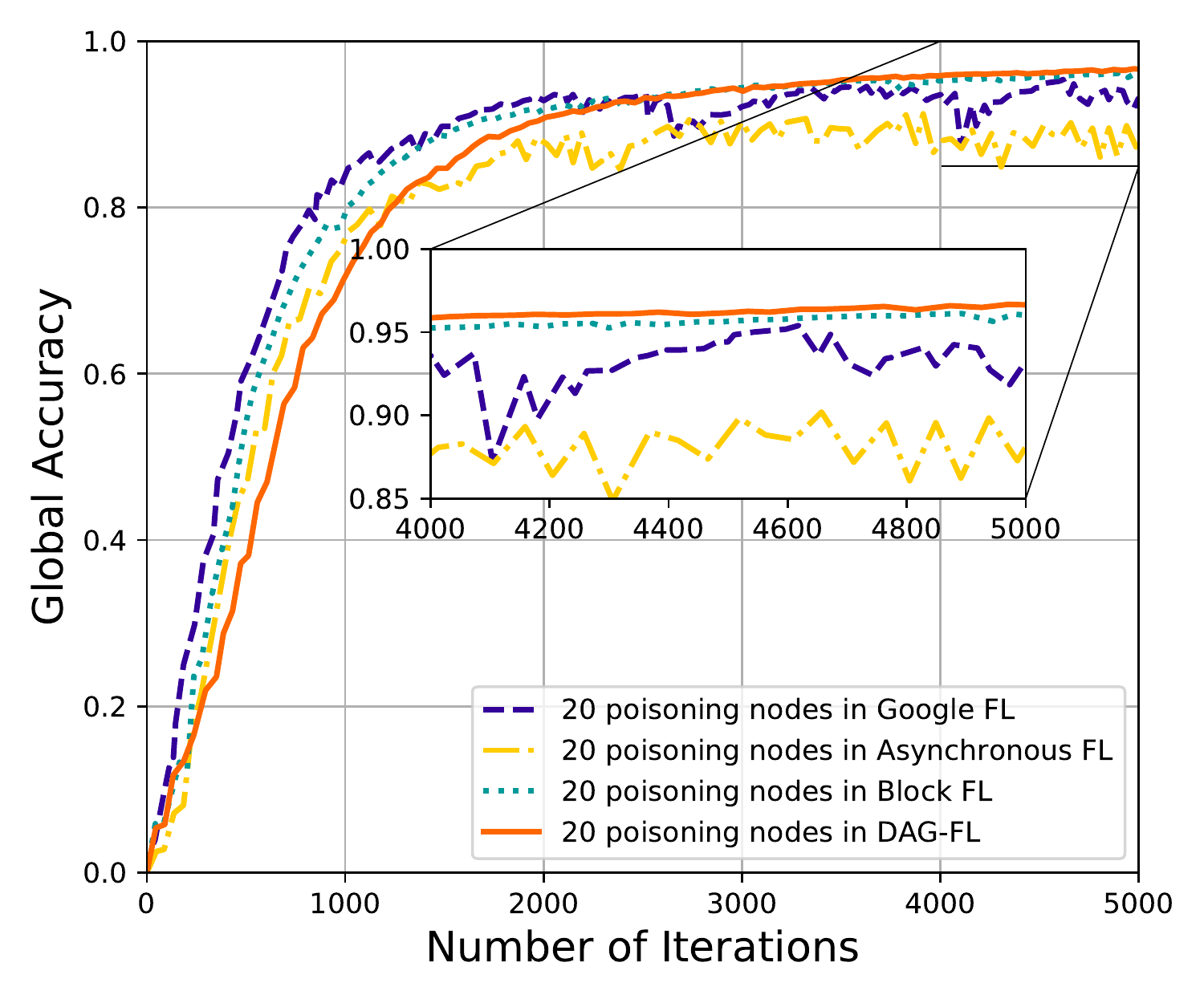}
		\caption{The accuracy of different FL systems with poisoning nodes for CNN task.}
		\label{cnn_i}
	\end{figure}

\begin{figure}[!t]
		\centering
		\includegraphics[width=3in]{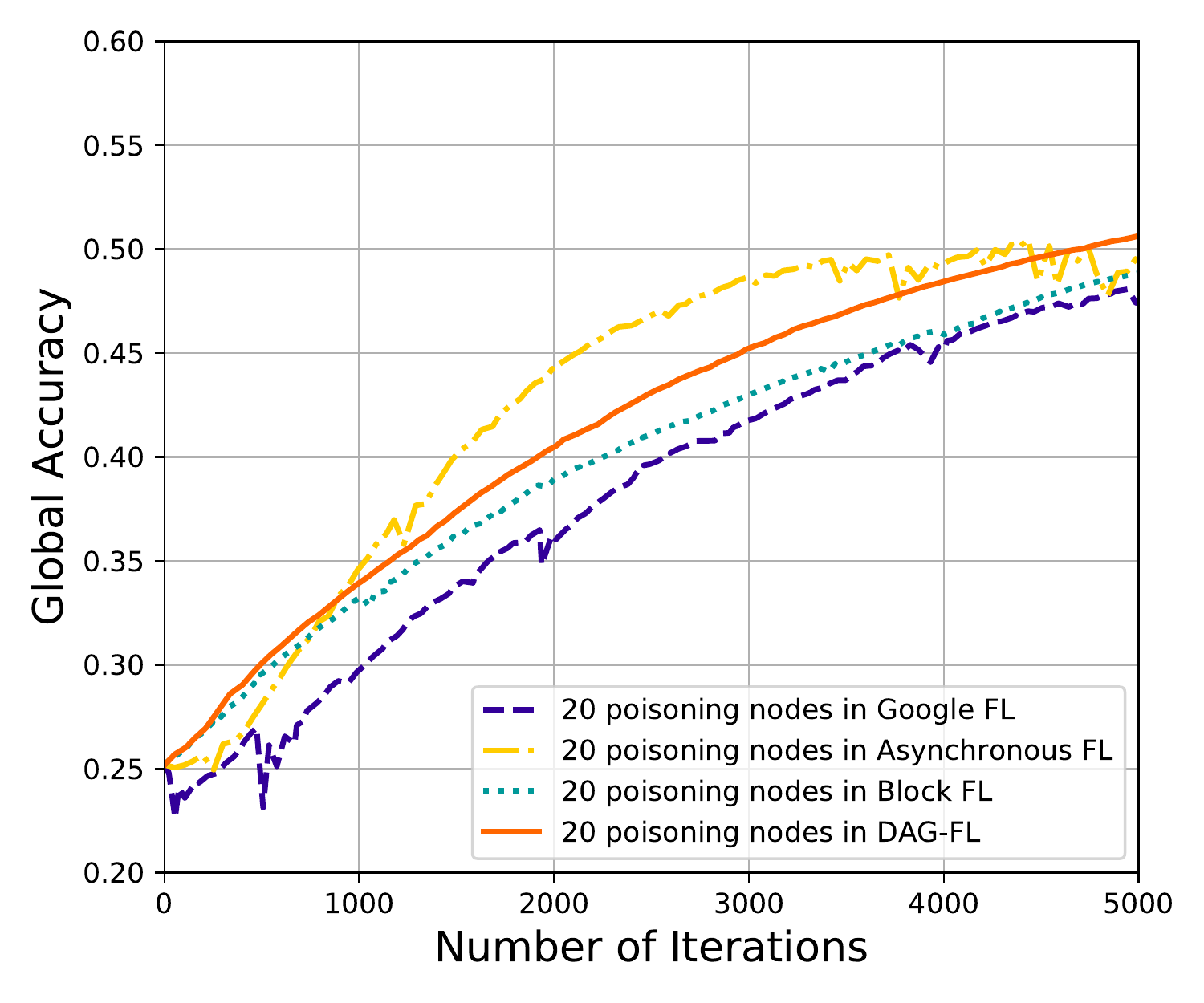}
		\caption{The accuracy of different FL systems with poisoning nodes for LSTM task.}
		\label{lstm_i}
	\end{figure}

 We demonstrate the accuracy curves of different FLs with 20 percent poisoning nodes in Fig. \ref{cnn_i} for CNN task and Fig. \ref{lstm_i} for LSTM task. We can see that accuracy of Google FL and Asychronous FL significantly reduces when poisoning nodes are involved, this is due to the fact that Google FL and Asychronous FL are not capable of distinguishing poisoning nodes. In contrast, these two FLs can only use the average algorithm to dilute the harmful model parameters uploaded by poisoning nodes, so as to achieve the purpose of mitigation. Since there are multiple miners to verify the model parameters uploaded by the associated nodes, the accuracy of Block FL does not decline much when poisoning nodes participate. DAG-FL performs best among the four FLs when poisoning nodes are involved in both CNN and LSTM tasks. The reason is that in the process of FL of DAG-FL, each node will validate the published transactions through the unique voting consensus mechanism. As a result, the transactions published by poisoning nodes will be isolated on DAG with the progress of FL, so as to mitigate the impact of poisoning nodes.

	\par
\begin{figure}[!t]
		\centering
		\includegraphics[width=3in]{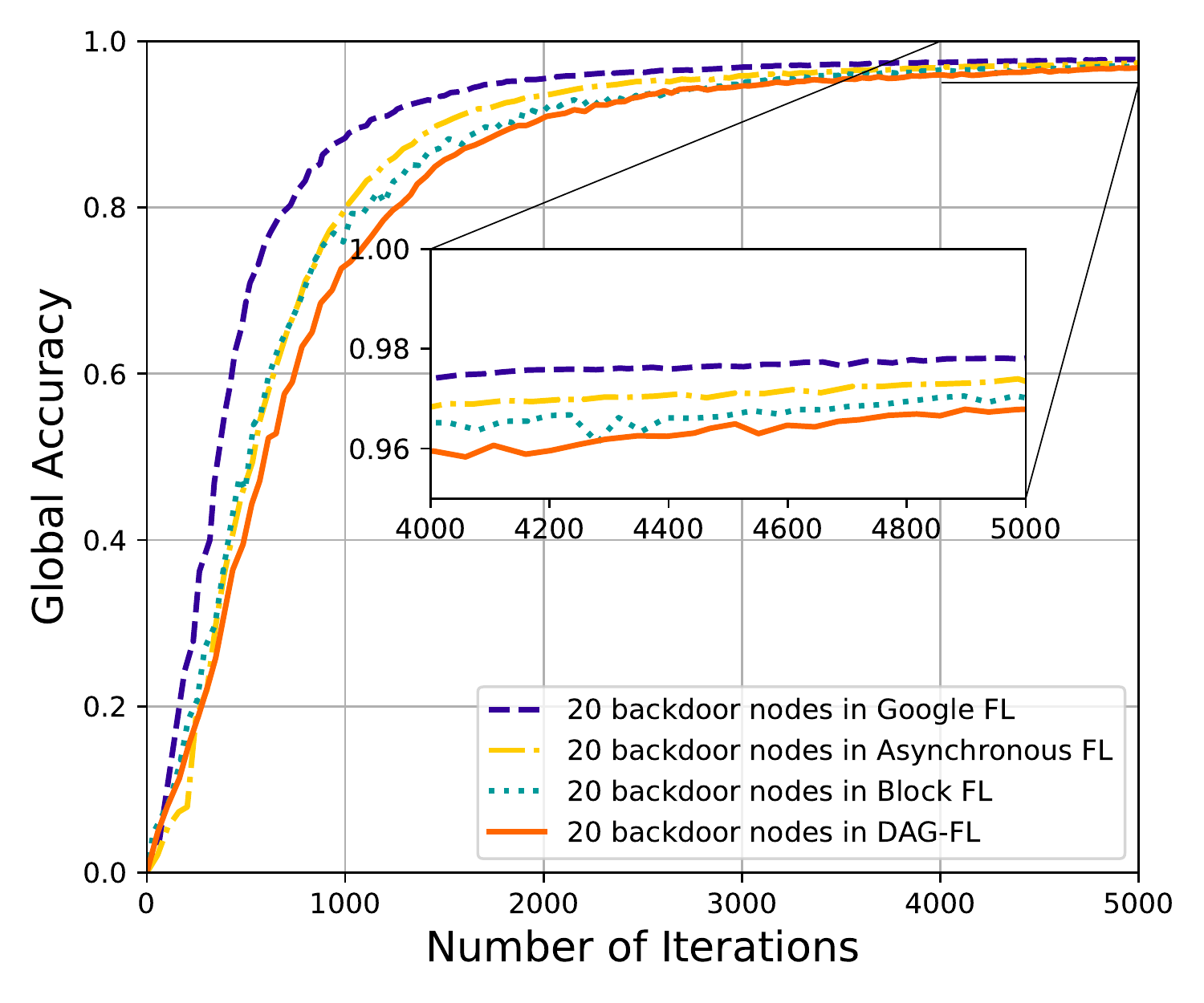}
		\caption{The accuracy of different FL systems with backdoor nodes for CNN task.}
		\label{cnn_t}
	\end{figure}

 Figure \ref{cnn_t} displays the accuracy curves of different FLs with 20 percent backdoor nodes in CNN task. It can be seen that all the four FLs have good convergence rates and accuracy similar to running in the ideal case. Block FL has a higher detection rate for models uploaded by backdoor nodes than DAG-FL because miners in Block FL use the whole test data set to validate models. This causes Block FL to perform better than DAG-FL in the first 5000 iterations when backdoor nodes are concerned. At 5000 iterations, the success rates of targeted attacks are computed for the four FLs as shown in Table III. The immunity of DAG-FL to targeted attacks is sensitive to the number of backdoor nodes. DAG-FL can well resist targeted attacks with an attack success rate of 0.006 when only 5 backdoor nodes are involved. When 20 backdoor nodes are involved, DAG-FL and Block FL perform similar immunity to targeted attacks, while Google FL and Block FL are completely captured with attack success rates above 0.9.

	\begin{table}[]
		\label{tar_table}
		\centering
		\caption{Attack success rates}
		\begin{tabular}{|c|c|c|}
			\hline
			System                  & Backdoor nodes & Attack success rate \\ \hline
			\multirow{3}{*}{DAG-FL} & 5              & 0.0060              \\ \cline{2-3} 
			& 10             & 0.3558              \\ \cline{2-3} 
			& 20             & 0.6243              \\ \hline
			Block FL                & 20             & 0.6193              \\ \hline
			Google FL               & 20             & 0.9166              \\ \hline
			Asynchronous FL         & 20             & 0.9211              \\ \hline
		\end{tabular}
	\end{table}

	\subsubsection{Detection of abnormal nodes}
	In the process of DAG-FL, we regard the transactions with less than or equal to $m$ approvals as isolated transactions which make no contribution to the FL, otherwise the transactions that exceed $m$ approvals can contribute. The contribution rate for a node in DAG-FL is the ratio of the number of contributing transactions to the number of all transactions published by the node. By comparing the contribution rates of nodes in DAG-FL, we can evaluate the ability of DAG-FL to do anomaly detection during the FL process. Table IV shows the contribution rates of DAG-FL in different scenarios with abnormal nodes after running for 10000s, in which $r$ is the average contribution rate of all nodes and $r_0$ is the average contribution rate of abnormal nodes. 
	
	For both CNN and LSTM tasks, it seems that lazy nodes cannot be distinguished in DAG-FL by comparing contribution rates when setting $m$ as 0. Through our analysis, the reason is that though the transactions published by lazy nodes cause worse staleness, most of transactions are not isolated due to less approvals. By setting $m=1$, DAG-FL seems to have good ability of anomaly detection when lazy nodes are few.
	
	For poisoning nodes, by comparing the contribution rates for both CNN and LSTM tasks, we draw a conclusion that DAG-FL can easily distinguish the poisoning nodes during the FL process. However, with the increase of poisoning nodes, the average contribution rate of them increases accordingly, which means that the ability of poisoning nodes detection of DAG-FL is degraded.
	
	For backdoor nodes, we find that DAG-FL has an excellent ability to detect when the number is small. However, this ability falls off a cliff when the number of backdoor nodes increases. This is because the backdoor nodes can launch a joint attack with the number increasing. Backdoor nodes prefer to approve transactions published by other backdoor nodes, which reduces the possibility of transactions published by backdoor nodes of being isolated. Thus, this joint attack can slash the anomaly detection ability of DAG-FL and succeed in creating a back door in the final target model of FL.

	\begin{table}[]
		\label{approval}
		\centering
		\caption{Contribution rates in different scenarios}
		
		\begin{tabular}{|c|c|c|c|c|c|c|}
			\hline
			\ Task          & Abnormal type     & $m$       & Nodes & $r_0$ & $r$ & $r_0/r$ \\ \hline
			\multirow{18}{*}{CNN}  & \multirow{6}{*}{Lazy}      & \multirow{3}{*}{0} & 5              & 0.779          & 0.827        & 0.941            \\ \cline{4-7} 
			&                            &                    & 10             & 0.788          & 0.827        & 0.952            \\ \cline{4-7} 
			&                            &                    & 20             & 0.783          & 0.817        & 0.958            \\ \cline{3-7} 
			&                            & \multirow{3}{*}{1} & 5              & 0.437          & 0.560        & 0.780            \\ \cline{4-7} 
			&                            &                    & 10             & 0.461          & 0.556        & 0.830            \\ \cline{4-7} 
			&                            &                    & 20             & 0.520          & 0.556        & 0.936            \\ \cline{2-7} 
			& \multirow{6}{*}{Poisoning} & \multirow{3}{*}{0} & 5              & 0.178          & 0.809        & 0.221            \\ \cline{4-7} 
			&                            &                    & 10             & 0.352          & 0.796        & 0.442            \\ \cline{4-7} 
			&                            &                    & 20             & 0.615          & 0.822        & 0.749            \\ \cline{3-7} 
			&                            & \multirow{3}{*}{1} & 5              & 0.033          & 0.555        & 0.059            \\ \cline{4-7} 
			&                            &                    & 10             & 0.074          & 0.551        & 0.135            \\ \cline{4-7} 
			&                            &                    & 20             & 0.277          & 0.562        & 0.493            \\ \cline{2-7} 
			& \multirow{6}{*}{Backdoor}  & \multirow{3}{*}{0} & 5              & 0.214          & 0.808        & 0.264            \\ \cline{4-7} 
			&                            &                    & 10             & 0.507          & 0.814        & 0.623            \\ \cline{4-7} 
			&                            &                    & 20             & 0.771          & 0.830        & 0.928            \\ \cline{3-7} 
			&                            & \multirow{3}{*}{1} & 5              & 0.050          & 0.568        & 0.089            \\ \cline{4-7} 
			&                            &                    & 10             & 0.220          & 0.553        & 0.398            \\ \cline{4-7} 
			&                            &                    & 20             & 0.461          & 0.561        & 0.821            \\ \hline
			\multirow{12}{*}{LSTM} & \multirow{6}{*}{Lazy}      & \multirow{3}{*}{0} & 5              & 0.742          & 0.827        & 0.897            \\ \cline{4-7} 
			&                            &                    & 10             & 0.777          & 0.829        & 0.937            \\ \cline{4-7} 
			&                            &                    & 20             & 0.793          & 0.830        & 0.955            \\ \cline{3-7} 
			&                            & \multirow{3}{*}{1} & 5              & 0.406          & 0.560        & 0.726            \\ \cline{4-7} 
			&                            &                    & 10             & 0.455          & 0.560        & 0.812            \\ \cline{4-7} 
			&                            &                    & 20             & 0.519          & 0.569        & 0.913            \\ \cline{2-7} 
			& \multirow{6}{*}{Poisoning} & \multirow{3}{*}{0} & 5              & 0.237          & 0.793        & 0.299            \\ \cline{4-7} 
			&                            &                    & 10             & 0.418          & 0.789        & 0.530            \\ \cline{4-7} 
			&                            &                    & 20             & 0.638          & 0.821        & 0.778            \\ \cline{3-7} 
			&                            & \multirow{3}{*}{1} & 5              & 0.029          & 0.557        & 0.051            \\ \cline{4-7} 
			&                            &                    & 10             & 0.090          & 0.546        & 0.166            \\ \cline{4-7} 
			&                            &                    & 20             & 0.248          & 0.553        & 0.448            \\ \hline
		\end{tabular}
	\end{table}

	\subsection{Testbed}
\begin{figure*}[!t]
		\centering
		\includegraphics[width=6in]{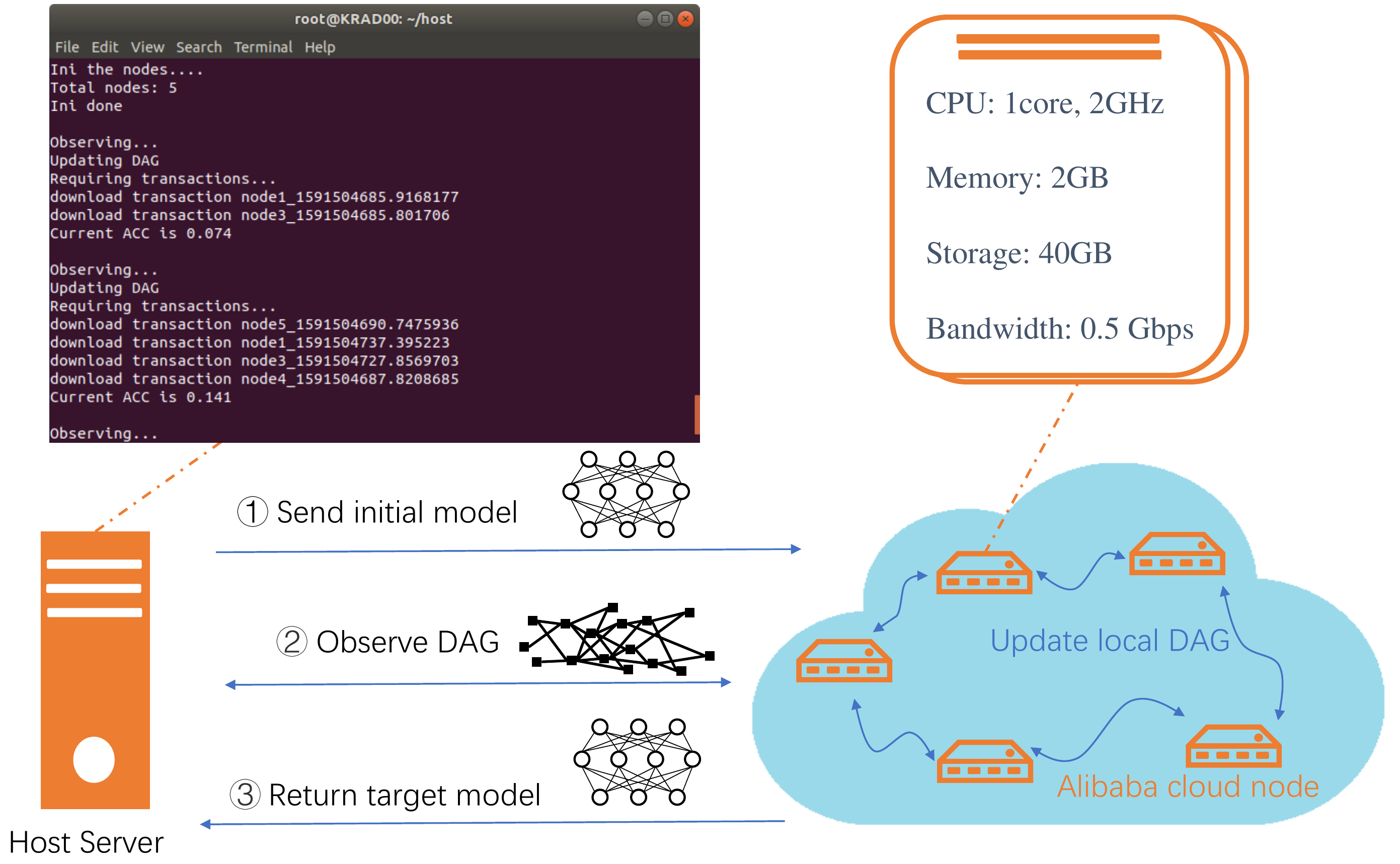}
		\caption{Testbed implementation}
		\label{i_s}
	\end{figure*}

	In order to illustrate the feasibility of DAG-FL, a practical application program is implemented to achieve DAG-FL on a real testbed with 5 cloud nodes and a host server as shown in Fig. \ref{i_s}. In the testbed, 5 nodes supported by the Alibaba Cloud Computing Company act as mobile devices with similar computing capacity and high network bandwidth, and a host server acts as the external agent $E$ with a host program running \textit{DAG-FL controlling} algorithm. 
	
	\begin{figure}[!t]
		\centering
		\includegraphics[width=3in]{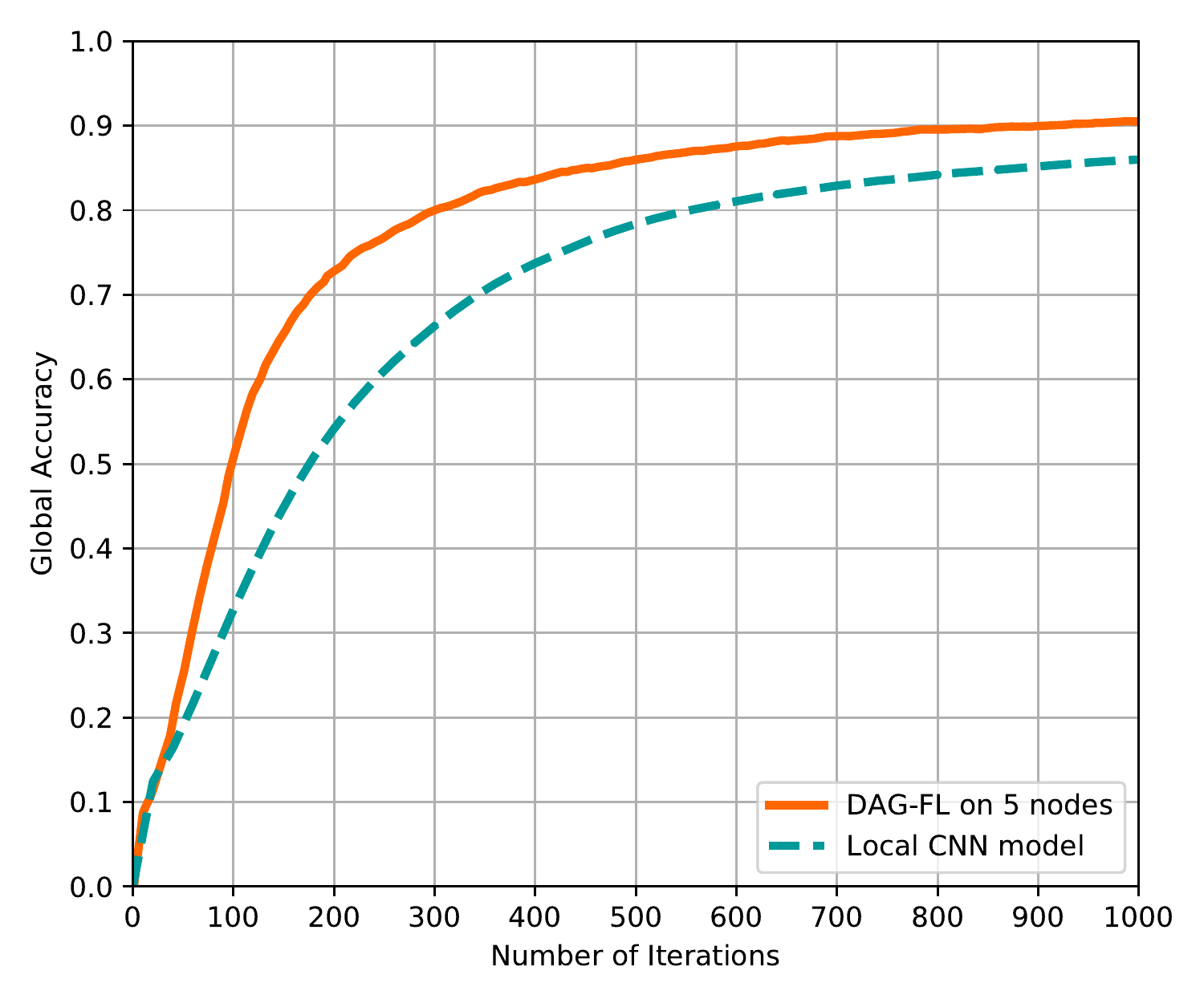}
		\caption{The accuracy of DAG-FL in real implementation.}
		\label{i_c}
	\end{figure}
	
	The goal of the host server includes: 1) initialize all other five participating nodes; 2) maintain a local DAG; 3) regularly compute a global model from the local DAG to observe whether the machine learning model co-construction is completed. If host server finds that FL is completed, it will send a termination signal to all nodes participated in DAG-FL. For other 5 nodes, they execute the client program to run \textit{DAG-FL updating} algorithm, which can maintain the local DAG by communicating with adjacent nodes periodically and actively participate in iterations of FL when nodes are in idle state. In Fig. \ref{i_c}, compared with training CNN on a single node, our DAG-FL deploying on 5 nodes has almost the same model convergence rate as the local CNN model in the first 50 iterations. This is because that temporarily constructed global models differ a lot from each other in the early stage. In the middle stage, different global models tend to integrate with each other, resulting in a sharp increase in the convergence rate of DAG-FL. Moreover, DAG-FL on 5 nodes have more training data to build CNN model, so as to achieve higher accuracy. In such an implementation experiment, we can conclude that DAG-FL can operate effectively in real situation, and achieve a good performance when building a CNN model.

	\section{Discussion}
	In this paper, we propose a DAG-FL framework to solve the problems of device asynchrony and anomaly detection, which is the first attempt to systematically apply DAG ledger technology to FL. In this section, we will discuss how to extend DAG-FL and highlight several problems to be solved in DAG-FL as future work in three aspects.
	
	\subsection{Model Validation}
	The model validation method of DAG-FL consensus in this paper is using a small test set to compute the prediction accuracy directly, which may not be applicable in certain federated learning scenarios \cite{shokri2015privacy}. We believe that using a more advanced model validation method can enhance the immune ability of DAG-FL to abnormal nodes, reduce the delay of an iteration of DAG-FL, and improve the overall system efficiency. For example, use a pre-trained autoencoder model like \cite{li2019abnormal} to detect abnormal transactions when validating during an iteration of DAG-FL.
	
	
	\subsection{Credit Evaluation}
	In DAG-FL, abnormal nodes can be detected by computing the contribution rate. Abnormal nodes with the low contribution in DAG-FL should be punished during the process of FL. In order to timely avoid the harm of abnormal nodes, it is feasible to design a confidence algorithm as a criterion for tips selection based on the contribution rate of each node, in which the tips generated by normal nodes could be selected with a higher probability. In contrast, abnormal nodes with low credit could be identified, and thus the published FL results as tips would be selected for validation rarely or orphaned finally. In this way, abnormal nodes are punished to be isolated, and normal nodes can get more potential rewards.
	
	\subsection{Weighted Aggregation}
	In this paper, global models are constructed by tips using the \textit{FederatedAveraging} algorithm with the same weight coefficient. However, when constructing a global model to train in an iteration of DAG-FL, even tips published by normal nodes are not equally important under an asynchronous environment, where some tips containing local models with better staleness and prediction accuracy are better due to the capability of the node. Therefore, a weight aggregation algorithm like \cite{kim2019blockchain} to assign a larger weight coefficient for the higher quality of the local model when constructing the global model. In this way, the constructed global model can better reflect the real progress of FL at the current time to improve the training efficiency of the whole FL and get a better target model finally.

	\section{Related Work}
	Currently, researches on FL have been received considerable attention to deal with the challenges of security and privacy. In this section, we review some state-of-art works of FL in four primary aspects: synchronous FL, pseudo-asynchronous FL, asynchronous FL, and blockchained FL.
	
	\subsection{Synchronous Federated Learning}
	Google FL \cite{mcmahan2017communication} and other similar FL systems [\citen{hard2018federated}, \citen{smith2017federated}, \citen{bonawitz2019towards}] are based on synchronous updating mechanism for federated learning. These systems usually allow central server to maintain a unique global model and assign FL tasks to some nodes that are in idle state in each round of FL. Before beginning another new round, the central server should wait for every selected nodes to complete their assigned FL tasks. This synchronous updating method promises a simple structure for FL and easy to implement, but errors on any node can block synchronous FL, resulting in poor tolerance of fault. In addition, the central server needs to upload and download models frequently, and is also responsible for maintaining entire network structure of FL. Thus, the scalability of synchronous FL is limited by the communication bandwidth and computing capacity of the central server \cite{li2020federated}.
	
	
	

	\subsection{Pseudo-asynchronous Federated Learning}
	Based on synchronous FL, researchers construct the pseudo-asynchronous FL to allow nodes doing FL iterations freely. Nodes in a pseudo-asynchronous FL system can download the global model from the central server at any time, and immediately upload their trained local models to a cache. The central server regularly updates its global model by aggregating trained local models from the latest global models in the cache. In order to improve the system efficiency, the interval between the global model updating should be set small, and bottleneck nodes usually cannot upload their trained local models in time. This causes meaningful data on bottleneck nodes are hard to utilize, which violates the original intention of making full use of the data on each node in FL. By redesigning nodes selection and global model aggregation methods, the authors in \cite{9093123} have established a pseudo-asynchronous FL system that can utilize data on bottleneck nodes. The central server in \cite{9093123} selects the local models trained from both the latest and outdated global models for aggregation when updating the global model. However, the aggregation method and the interval between the global model updating should be well designed for different FL scenarios.

	\subsection{Asynchronous Federated Learning}
	The authors in \cite{li2020federated} demonstrate that asynchronous scheme is an attractive approach to mitigate stragglers in heterogeneous environments. As FL is more likely to be deployed in mobile wireless networks which are usually heterogeneous environment in the future, some researches on designing the asynchronous FL systems have been investigated [\citen{sprague2018asynchronous}, \citen{chen2019asynchronous}]. In asynchronous FL, a node can download a global model from a central server and train a local model to upload in idle state at any time. In \cite{lian2018asynchronous}, the authors propose an asynchronous decentralized algorithm based on stochastic gradient decent. In a heterogeneous environment, the proposed algorithm is robust to build deep learning model like VGG \cite{jaderberg2015spatial} and ResNet-20 \cite{he2016identity}, and can achieve efficient communications with the best-possible convergence rate. Based on the work of \cite{lian2018asynchronous}, the authors in \cite{xie2019asynchronous} propose an asynchronous FL algorithm with an aim of weighting different local models of staleness to update global model. Meanwhile, considering both strongly and non-strongly convex problems, this work also proves that the asynchronous updating method can reach near-linear convergence. 

	\subsection{Blockchained Federeated Learning}
	It is noted that FL and blockchain technology are both deployed in a distributed network, many researchers attempt to combine the advantages of them and have proposed some influential blockchained FL systems [\citen{9163027}, \citen{salah2019blockchain}, \citen{marr2018artificial}]. Combining blockchain and FL, the authors in \cite{weng2019deepchain} propose a DeepChain architecture. This work focuses on solving the problems of privacy and auditability by storing the local machine learning model in blockchain and regarding these models as transactions, which uses the security and immunity of blockchain to prevent nodes without permission from accessing information stored in transactions. Similarly, in order to store sensitive and personal data in a diskless environment, research like \cite{lu2019blockchain} adopts blockchain to form a data sharing platform which is beneficial for data privacy and security. On this basis, the authors in \cite{9210138} utilize a sharding-based blockchain protocol to well protect the convergence of FL when large-scale nodes are involved. Furthermore, researchers in \cite{8998397} also propose a sharding-based FL system by applying a two-layer blockchain framework.

	In recent works [\citen{kim2019blockchained}, \citen{majeed2019flchain}], researchers are interested in a deep combination of blockchain and FL and get more suitable models for on-device scenarios or other. Both of the works use the miner in blockchain to complete the task of the central server in traditional FL. The authors in \cite{kim2019blockchained} design a Block FL with multiple miners to coordinate FL tasks and maintain the global model, and the procedure can be described as follows: 1) The node downloads the global model from its associated miner and get a trained local model. 2) The trained local model will be uploaded to the associated miner as a transaction. 3) The miner will confirm the validity of the uploaded transaction, and the confirmed transactions will be stored in the candidate block of the miner. Once candidate blocks collect enough transactions or wait for a certain time, all miners enter the consensus stage together and run PoW to compete for a winner to publish its own candidate block on the blockchain. Besides, miners are capable of allocating rewards when publishing their blocks on blockchain to motivate mobile devices to participate in FL. Followed up the work \cite{kim2019blockchained}, the authors in \cite{majeed2019flchain} propose a FLchain architecture that considers MEC servers as miners to solve problems brought by central servers in traditional FLs. The FL systems in these two works are more like the pseudo-asynchronous ones, in which nodes are freed after uploading local trained models to miners. Compared with only using blockchain to store and transmit information, works like [\citen{kim2019blockchained}, \citen{majeed2019flchain}] provide a feasible way to achieve efficient fusion of blockchain and FL in terms of data security and privacy.
	
	However, these previous works still remain issues to be addressed on both blockchain and FL. On the one hand, these works insist on the PoX consensus mechanism, which naturally leads to the pseudo-asynchronous framework that cannot meet the device asynchrony in on-device FL. Besides, the introduced miners would generate extra delay and resource consumption on consensus but not for FL at all. On the other hand, these works rely on miners to validate uploaded local models stored in transactions, which might cause miner's dilemma due to the conflict of interest between blockchain and FL. As the transaction validation is executed on miners, the target model might be trained towards the expected direction of miners instead of nodes, resulting in the problem of local optima.

	To this end, for asynchronous bookkeeping and low resource consumption of DAG ledger technology, this paper proposes DAG-FL introducing DAG empowers asynchronous FL framework. Through the workload assign of consensus and model validation on each node in a decentralized manner, miners and mining are not needed anymore. By combining the voting mechanism of DAG ledger technology with the process of local model validation in FL, DAG-FL can detect abnormal nodes and mitigate their impact autonomously to guarantee security. Meanwhile, as the model validation is operated on each node without professional miners, the target model of FL would be trained under the observation of all nodes avoiding the training deviation for any reason.

	\section{Conclusion}
	In this paper, we proposed a DAG empowered FL system named as DAG-FL, which combines DAG ledger technology to overcome the problems of mobile device asynchrony and abnormal nodes to complete the model co-construction task of FL efficiently. DAG-FL is constructed by a three-layer asynchronous architecture including FL, DAG, and application layers taking the responsibilities of model training, communicating and observing. In this manner, DAG layer could provide blockchain as a service (BaaS) establishing \textit{DAG-FL controlling} and \textit{DAG-FL updating} algorithms for operation process of DAG-FL. Meanwhile, to maintain the stability and efficiency of DAG-FL, we discussed two deployment details explaining how to keep the reasonable number of tips on DAG and how to address the staleness of FL results. Finally, experimental results showed that DAG-FL has the higher system efficiency and a better target model with device asynchrony compared with other three benchmark FL systems. Meanwhile, DAG-FL is also proved to be insensitive to the impact of abnormal nodes and has the ability of anomaly detection. Moreover, the results on the testbed where DAG-FL can also achieve the high stability and efficiency as expected. To provide some open discussion issues to extend DAG-FL, we highlighted the concerns on model validation, credit evaluation and weighted aggregation as the possible future direction.

	\bibliographystyle{IEEEtran}      
	\bibliography{refrence}                        

\end{document}